\documentclass[prc,notitlepage, twocolumn,
showpacs,floatfix,nofootinbib,preprintnumbers,superscriptaddress,amsmath,amssymb,aps,longbibliography,10pt]{revtex4-2}

\usepackage[dvipdfmx]{graphicx}
\usepackage{epsfig}
\usepackage{amsmath,amssymb,amsfonts}
\usepackage{bm}
\usepackage{color}
\usepackage{float}
\usepackage{dcolumn} 
\usepackage{enumerate}
\usepackage{multirow}
\usepackage{threeparttable}
\usepackage{siunitx}
\usepackage{physics}
\usepackage{mathrsfs}
\usepackage[colorlinks=true,linkcolor=blue,citecolor=blue]{hyperref}

\graphicspath{{fig/}}
\allowdisplaybreaks

\begin{document}

\title{
Moment of inertia for pair rotation:\\
Interplay between order parameter and shell structure
}
\author{Chisato Ruike}
\email{ruike@nucl.ph.tsukuba.ac.jp}
\affiliation{
Graduate School of Science and Technology, University of Tsukuba, Tsukuba 305-8571,  Japan}
\author{Nobuo Hinohara}
\email{hinohara@nucl.ph.tsukuba.ac.jp}
\affiliation{
 Center for Computational Sciences, University of Tsukuba, Tsukuba, Ibaraki 305-8577, Japan
}
\affiliation{
 Faculty of Pure and Applied Sciences, University of Tsukuba, Tsukuba, Ibaraki 305-8571, Japan
}
\affiliation{Facility for Rare Isotope Beams, 
              Michigan State University, East Lansing, Michigan 48824, USA}
\author{Takashi Nakatsukasa}
\email{nakatsukasa@nucl.ph.tsukuba.ac.jp}
\affiliation{
 Center for Computational Sciences, University of Tsukuba, Tsukuba, Ibaraki 305-8577, Japan
}
\affiliation{
 Faculty of Pure and Applied Sciences, University of Tsukuba, Tsukuba, Ibaraki 305-8571, Japan
}
\affiliation{RIKEN Nishina Center for Accelerator-Based Science, Wako, Saitama 351-0198, Japan}

\date{\today}

\begin{abstract}
\begin{description}
\item[Background] Pair condensation in finite nuclei generates a collective motion known as pair rotation. The moment of inertia of pair rotation (P-MoI) has been used as an indicator of pair condensation.
\item[Purpose] We aim to elucidate the fundamental properties of the P-MoI, particularly its dependence on the particle number and the order parameter.
\item[Method] The P-MoI 
 was evaluated using the Bardeen-Cooper-Schrieffer (BCS) calculations with a monopole pairing Hamiltonian and Skyrme density functional theory calculations with different density-dependent pairing energy density functionals.
\item[Results] In open-shell nuclei, a negative correlation was found between the P-MoI and the pair amplitude, which is the order parameter for the transition from the normal phase to the superconducting phase. 
Analysis based on the decomposition of the P-MoI into the orbital contributions within the BCS approximation shows that its 
orbital dependence is very similar to that of the pairing gap.
\item[Conclusions] The P-MoI reflects the influence of both the level density near the Fermi energy and the pair amplitude.
\end{description}
\end{abstract}

\maketitle

\section{Introduction}
Pairing correlation plays an essential role in the nuclear structure \cite{Dean2003,Brink-Broglia}.
The most prominent pairing correlation is that
two like-nucleons form a spin-singlet pair with zero total angular momentum ($^1S_0$ pairing).
The existence of the $^1S_0$ pairing in nuclei has typically been evidenced by systematic differences between the properties of nuclei with even and odd numbers of neutrons $N$ (protons $Z$).
The masses of nuclei with even $N$ ($Z$) are larger than those of odd-$N$ (odd-$Z$) nuclei.
This is known as the odd-even mass staggering (OES).
Other signatures of the pairing correlation include
enhanced cross sections in the pair-transfer reactions
\cite{Potel2011,Potel2013,PhysRevC.87.054321,PhysRevC.92.064602},
reduction of the moment of inertia (MoI) for deformed nuclei
\cite{RS80},
and difference in the level densities
between even and odd systems \cite{BM75}.

\begin{figure}[ht]
\begin{center}
\includegraphics[width=70mm]{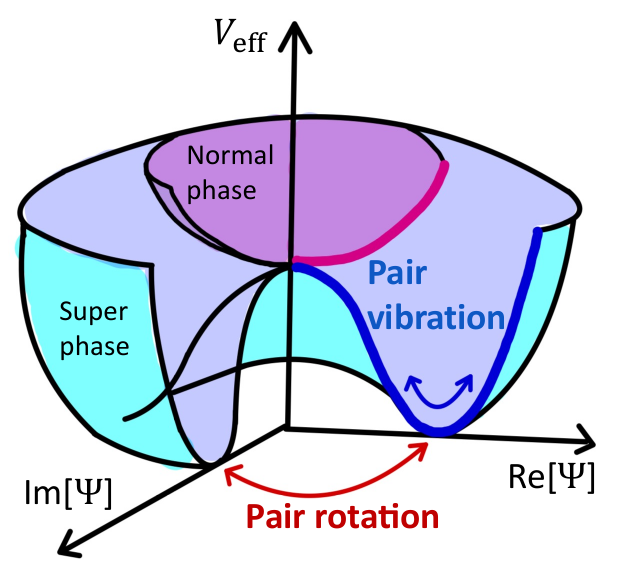}
\caption{Schematic figure of the effective potential as a function of the order parameters $\Psi$.
The pairing rotation corresponds to the rotational degree of freedom
in the phase of $\Psi$.}
    \label{fig:potential}
    \end{center}
\end{figure}

The nuclear pairing correlation can be treated with the Bardeen-Cooper-Schrieffer (BCS) theory \cite{BMP58},
analogously to the superconductor in metals,
introducing a one-body pair potential that spontaneously breaks the rotational symmetry in the gauge space.
The BCS wave function has a specific orientation in the gauge space that corresponds to a certain body-fixed frame.
Usually, we fix the orientation by setting the $(U,V)$ coefficients in the BCS wave function to be real numbers
\cite{Broglia2000}.
This spontaneous symmetry breaking is illustrated in Fig.~\ref{fig:potential}
with the effective potential as a function of the relevant collective variables (order parameters) $\Psi$ which is complex.
When the gauge symmetry is spontaneously broken,
the location of the potential minimum changes from the origin to the bottom of the valley with a nonzero value of the order parameters $\Psi\neq 0$.
Although the states at the bottom of the effective potential are
infinitely degenerate,
an arbitrary choice of the state breaks the global U(1) gauge symmetry.
The pair rotation emerges as a collective motion along the valley
~\cite{PhysRev.112.1900,PhysRev.117.648,INC_19_154, Papenbrock2022},
known as the Nambu-Goldstone mode,
which restores the broken symmetry.
The pair rotation is visible in the rotational spectra of the ground-state energy with different particle numbers $N$ ($Z$), analogous to the rotational spectra with different values of angular momentum.
The MoI for the pair rotation,
which is hereafter abbreviated as P-MoI, 
is the inverse of the second derivative of the ground-state energy with respect to the particle number $N$ ($Z$).

In the nuclear mean-field theory or the density functional theory (DFT), one normally identifies the theoretical pairing gap with the magnitude of OES.
However, there are ambiguities in the 
definitions of both the pairing gap and the experimental OES,
which makes precise comparison of the theoretical and experimental values difficult \cite{Bender2000}. 
The computation of the ground-state energies of the odd-mass nuclei
may directly provide the OES.
However, since the time-odd mean fields of the energy density functional (EDF) are not well constrained at present \cite{Bertsch2009,Schunck2010},
the calculated values of the OES may not be so reliable.

The P-MoI serves as an ideal observable free from the issues described above.
It can be calculated from the ground-state energies of three neighboring even-even nuclei, where the time-odd fields are absent in the EDF calculations.
The connection between the P-MoI and the two-nucleon separation energy has been discussed before \cite{ZPA250_155,Krappe1975}.
The P-MoI may be a useful quantity to determine the parameters and the density dependence of the pairing EDF  \cite{Hinohara2016,Hinohara2018,Reinhard2017,Reinhard_2024,Kouno2021}.
The P-MoI should have a strong correlation with the property of the pair condensation in the ground state. 
This paper aims to 
clarify the fundamental property of the P-MoI,
in particular, similarities, differences, and correlations between the P-MoI and the magnitude of the order parameter.

Similarities between rotation in the real space and that in the gauge space have been discussed in the past \cite{Brink-Broglia,Broglia2000}.
The rotational spectra in deformed nuclei
have been extensively studied,
for which the MoI ${\cal J}$
is believed to increase monotonically as a function of
the deformation parameter $\beta$.
The rigid rotor (irrotational flow) model predicts
the dependence on deformation as
${\cal J}\propto \beta$ ($\beta^2$)
\cite{BM75}.
At least for intrinsic states with equilibrium deformation,
the MoI ${\cal J}$ increases as $\beta$ increases\footnote{
It is worth noting that the MoI at non-equilibrium states is shown to be not necessarily a monotonic function of deformation
\cite{WN22}.
}.
However, the basic properties of the P-MoI remain unsettled.
In the special case of a single-level pair model
\cite{RS80},
which is analytically solvable,
the P-MoI does not depend on the number of particles but only on the pairing force strength.
We adopt a multi-level pair model to investigate
the P-MoI in detail, with a comparison of results
of realistic Skyrme EDF models.

This paper is organized as follows.
In Sec.~\ref{sec:prmoi}, the basic property of the P-MoI is summarized.
Section~\ref{sec:method} summarizes the adopted models to compute the ground states.
The analysis of the P-MoI is given in Sec.~\ref{sec:results}.
The conclusion is given in Sec.~\ref{sec:conclusion}.

\section{Moment of inertia for pair rotation (P-MoI) \label{sec:prmoi}}

Let us focus on the ground-state energies of even-even spherical proton-magic nuclei
where only the neutrons contribute to the pairing correlation.
The ground-state energies are expanded in the Taylor series as a function of the neutron number $N$ around a certain number $N_0$.
If the reference system at $N=N_0$ is superconducting, 
the ground-state energies of even-mass nuclei are well approximated by the second-order expansion with respect to $N-N_0$ as
\begin{align}
E_{0}(N)&\approx E_{0}(N_{0})+ \lambda(N_0)
(N-N_{0})
+\frac{(N-N_0)^2}{2{\mathscr{I}(N_0)}}.
\label{E0expand}
\end{align}
See Fig.~2 in Ref.~\cite{Broglia2000} for an example.
Here, $E_{0}(N)$ represents the ground-state energy of the nucleus with the neutron number $N$,
and $\lambda(N_0) = dE/dN|_{N=N_0}$
is the Fermi energy of the nucleus with the neutron number $N_0$.
The second-order term in Eq. (\ref{E0expand})
is referred to as the pairing rotational energy,
with the P-MoI given by $\mathscr{I}(N_0) = [d^2E/dN^2|_{N=N_0}]^{-1}$.
The pairing rotational energy has similarities to the rotational energy of two-dimensional systems
with the rotational energy $E_{\rm rot}=J^2/2{\cal J}$
with the angular momentum $J$ and the 
dynamical rotational MoI
${\cal J}\equiv dJ/d\omega = 
(d^2 E_{\rm rot}/d J^2)^{-1}$.  

In this paper, we evaluate the P-MoI using the three-point
finite-difference formula,
with even-$N$ nuclei unless otherwise noted.
\begin{align}
{\mathscr I}(N)
&=\biggr[\frac{E_{0}(N-2)-2E_{0}(N)+E_{0}(N+2)}{4}\biggr]^{-1}.
\label{eq:pairingMOIapr}
\end{align}
The inverse of the P-MoI evaluated in this way,
precisely speaking $4{\mathscr I}^{-1}$,
is equivalent to the two-nucleon empirical shell gap~\cite{Lunney2003}
\begin{align}
 \delta_{2n}(N) &= 
 S_{2n}(N) - S_{2n}(N+2) \nonumber \\
 &= E_0(N-2) - 2E_0(N) + E_0(N+2) ,
 \label{eq:DS2n}
\end{align}
where $S_{2n}$ is the two-neutron separation energy.

\section{Models \label{sec:method}}
We use two models to compute the ground-state energies
of even-even nuclei.
One is a pairing Hamiltonian with
phenomenological single-particle energies,
treated with the BCS approximation.
Another one is the Skyrme EDF with the volume, mixed, and surface types pairing EDFs,
which may validate the results of the pairing Hamiltonian
in realistic nuclei.

\subsection{Pairing Hamiltonian}
In this work, we study the pairing Hamiltonian
for the spherical nuclei
\begin{align}
\hat{H}
&=\sum_{j} e_{j}\sum_{m>0}(\hat{a}_{jm}^{\dagger}\hat{a}_{jm}+\hat{a}_{\widetilde{jm}}^{\dagger}\hat{a}_{\widetilde{jm}})
-G \hat{A}^\dag\hat{A},
\label{H}
\end{align}
where $e_j$ are the single-particle energies and
the index $j$ means the angular momentum.
In general, the index for the spherical single-particle energy
should represent a set of quantum numbers,
namely, the radial quantum number $n$, the orbital angular momentum $l$, and the total angular momentum $j$.
However, the single-particle energies adopted in the present study (cf. Sec.~\ref{sec:model_parameters})
can be specified by $j$ only.
The single-particle energies with different magnetic quantum numbers $m$ are $2\Omega_j$-fold degenerate with $\Omega_j\equiv j+1/2$.
The second term in the right-hand side of Eq. (\ref{H})
is the monopole-pairing interaction with the pairing force strength $G$.
The pair operators are defined as
\begin{equation}
\hat{A}^{\dag}\equiv 
\sum_j \sum_{m>0}
\hat{a}_{jm}^{\dagger}\hat{a}_{\widetilde{jm}}^{\dagger} ,
\quad\quad
\hat{A}=\left(\hat{A}^\dag\right)^\dag
\end{equation}
where 
$\hat{a}^\dag_{jm}$ creates a nucleon at a state $jm$ while $\hat{a}^\dag_{\widetilde{jm}}$
does at the time-reversal state $\widetilde{jm}$.
The summation $\sum_{m>0}$ means that the summation is taken over the positive magnetic quantum numbers $m>0$ only.

The ground state is assumed to be in the following BCS form
\begin{align}
\ket{{\rm BCS}}=\prod_{j,m >0}(U_{j}+V_{j}\hat{a}_{jm}^{\dagger}\hat{a}_{\widetilde{jm}}^{\dagger})\ket{0},\label{BCSwf}
\end{align}
where $U_j$ and $V_j$ are chosen to be real and positive.
$V_j^2$ is the occupation probability for the orbit $(jm)$, and $U_{j}^{2}$ is defined by $U_{j}^{2}=1-V_{j}^{2}$.
The BCS wave function is a superposition of states with different particle numbers, thus,
it is not an eigenstate of the particle-number operator.
$U_{j}$ and $V_{j}$ in the BCS wave functions are determined by solving the 
variational equation under the constraint on the particle number
\begin{align}
\delta\bra{\rm BCS}\hat{H}-\lambda \hat{\mathscr{N}}\ket{\rm BCS}=0 \label{variational}
\end{align}
with the number condition
\begin{align}
 \bra{\rm BCS}\hat{\mathscr{N}}\ket{\rm BCS} = N.
\end{align}
Here $\hat{\mathscr{N}}$ is the particle-number operator whose expectation value is constrained to a specific neutron number $N$, and the $\lambda$ is the Lagrange multiplier.
Using the degeneracy of the orbits $2\Omega_{j}$, 
the ground-state energy can be written as follows,
\begin{align}
E_{0}=2\sum_{j} e_{j}\Omega_{j} V_{j}^{2}-\frac{\Delta^{2}}{G}-G\sum_{j}\Omega_{j} V_{j}^{4}, \label{E0}
\end{align}
where $\Delta$ is the pairing gap
\begin{align}
\Delta&\equiv G\bra{\rm BCS}\hat{A}\ket{\rm BCS} = 
G\sum_{j}\Omega_{j} U_{j}V_{j}. \label{Delta}
\end{align}
The order parameter of the pair condensation 
is defined by the pair amplitude,
the expectation value of the pair operator $\bra{\rm BCS}\hat{A}\ket{\rm BCS}$.
In the present case,
it is equal to $\Delta/G$.

\subsection{Skyrme EDF}
We also perform DFT calculations using the Skyrme EDF as non-empirical realistic calculations.
For details, see e.g., Refs.~\cite{Stoitsov2005, Vautherin1972, Bender2003}.
The pairing potential for neutrons is given in the following form
\begin{align}
{\tilde{h}}_{n}(\textbf{r})=\frac{V_{0}}{2}\left[1-V_{1}\frac{\rho_0(\textbf{r})}{\rho_c}\right]\tilde{\rho}_{n}(\textbf{r}), 
\label{eq:HFBpairing}
\end{align}
where $\tilde{\rho}_n(\mathbf{r})$ denotes the neutron pair density, $\rho_0(\mathbf{r})$ is the isoscalar particle density,
and $\rho_c=0.16$ fm$^{-3}$ is the saturation density.
$V_0$ is the pairing strength, and $V_1$ controls the density dependence of the pairing potential.
In this paper, we use the volume ($V_1=0$), mixed ($V_1=0.5$), and surface ($V_1=1$) 
types of pairing.

\section{Results \label{sec:results}}

We study the neutron spin-singlet pairing in the proton-magic Ni, Sn, and Pb isotopes.

\subsection{Model parameters \label{sec:model_parameters}}

The BCS calculations are performed using the
pairing Hamiltonian (\ref{H}) for neutrons.
The single-particle levels are obtained with the modified oscillator potential \cite{Nilsson},
listed in Table~\ref{table:singleeng}.
The levels in Table~\ref{table:singleeng} also
indicates the employed single-particle model space for each isotope.
The pairing force strength is assumed to have a mass number dependence $G=G_0 A^{-1}$,
and the coefficient $G_0$ is
determined for each isotope 
so as to reproduce the 
experimental neutron OES in $^{68}$Ni, $^{116}$Sn, and $^{188}$Pb.
The obtained values are
$G=27.7/A$ MeV, $22.5/A$ MeV, and $18.9/A$ MeV, 
for Ni, Sn, and Pb isotopes, respectively.

The Skyrme DFT calculations are performed with the SLy4 EDF \cite{Chabanat1998} and the pairing EDF of Eq.~(\ref{eq:HFBpairing}).
We use the {\sc hfbtho} code \cite{Stoitsov2013}
by imposing the spherical symmetry.
The single-particle model space consists of the 20 harmonic oscillator major shells together with the 60 MeV cutoff to truncate the quasiparticle space.
The pairing strengths $V_0$ in Eq. (\ref{eq:HFBpairing}) are adjusted for each isotopic chain separately using the experimental OES of the same three nuclei,
$^{68}$Ni, $^{116}$Sn, and $^{188}$Pb.
They are, respectively,
$V_0=-226.5$, $-177.3$, and $-185.5$ MeV fm$^3$
for the volume-type pairing, and 
$V_0=-331.0$, $-267.9$, and $-290.0$ MeV fm$^3$ for the mixed-type pairing,
and $-$414.0, $-$431.0, and $-$530 MeV fm$^3$ for the surface-type pairing.

\begin{table}[H]
\begin{center}
\caption{Neutron single-particle energies of Ni, Sn, and Pb isotopes calculated with the modified oscillator potential \cite{Nilsson}.}
\begin{ruledtabular}
\sisetup{table-format=2.2}
\begin{tabular}{cScScS}
\multicolumn{2}{c}{Ni} & \multicolumn{2}{c}{Sn} &  \multicolumn{2}{c}{Pb} \\ \hline
$j$ &{$e_{j}$ (MeV)} &$j$  &{$e_{j}$ (MeV)} &$j$  &{$e_{j}$ (MeV)}\\ \hline
$2p_{3/2}$ &-9.94  &$2d_{5/2}$  &-10.362  & $2f_{7/2}$  &-11.392\\
$1f_{5/2}$ &-7.68  &$1g_{7/2}$  &-9.456   & $1h_{9/2}$  &-10.832\\ 
$2p_{1/2}$ &-7.23  &$3s_{1/2}$  &-7.808   & $1i_{13/2}$ &-9.355\\
$1g_{9/2}$ &-5.03  &$1h_{11/2}$ &-7.461   & $3p_{3/2}$  &-8.596\\ 
~          & ~     &$2d_{3/2}$  &-7.419   & $2f_{5/2}$  &-8.285\\ 
~          & ~     &         ~  & ~       & $3p_{1/2}$  &-7.264\\ 
\end{tabular}\label{table:singleeng}
\end{ruledtabular}
\end{center}
\end{table}

\subsection{P-MoI for Ni, Sn, and Pb isotopes}

\begin{figure}[ht]
\begin{center}
\includegraphics[width=80mm]{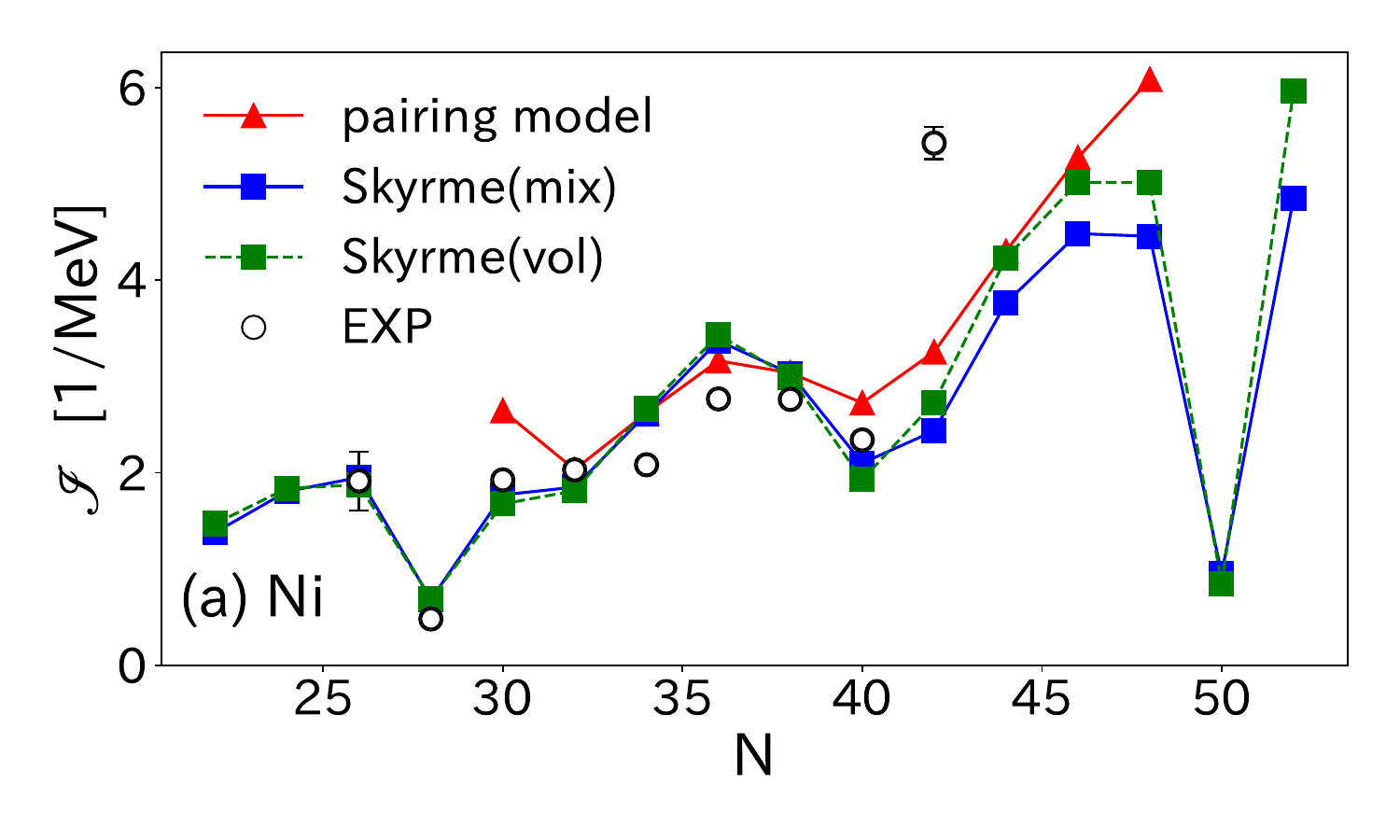} \\
\includegraphics[width=88mm]{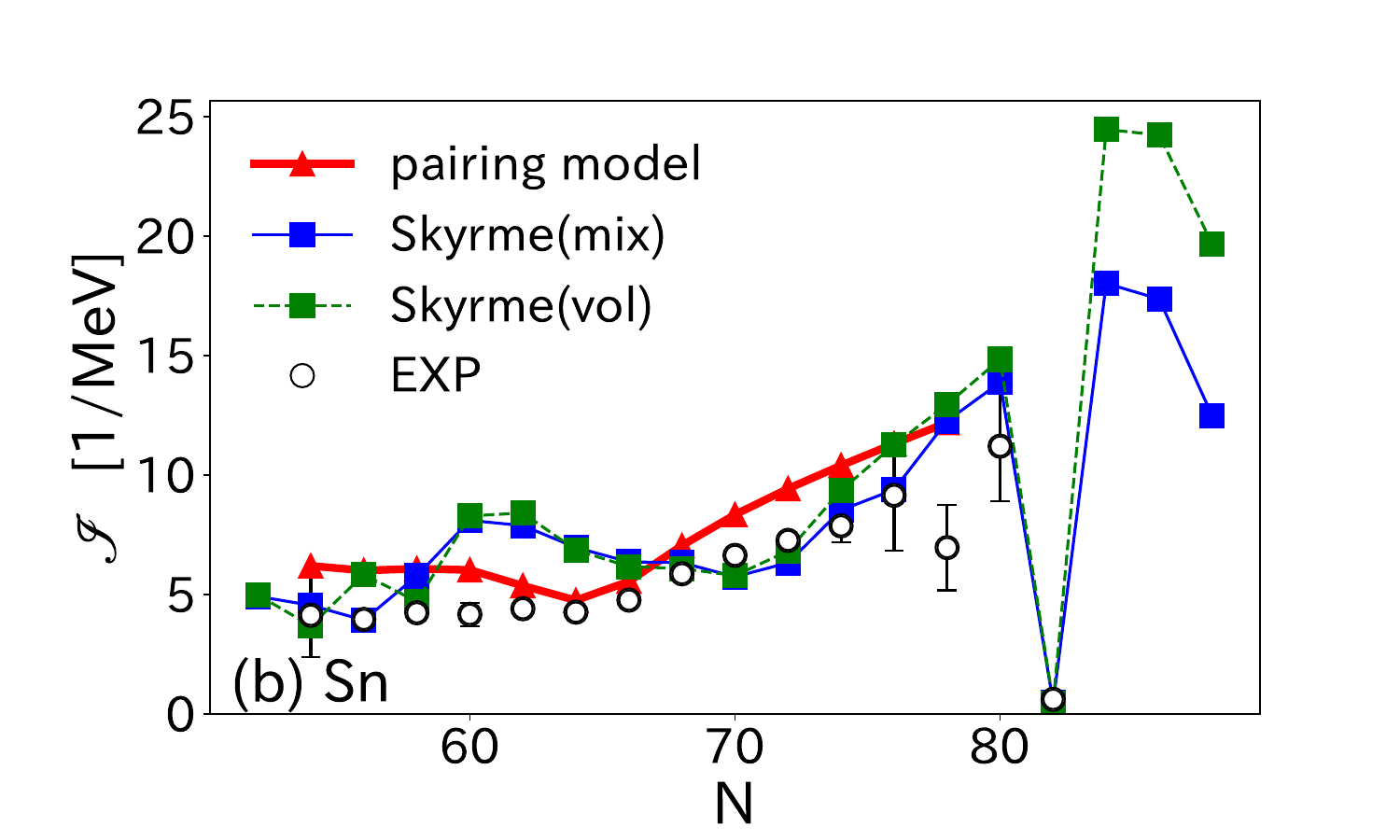} \\
\includegraphics[width=80mm]{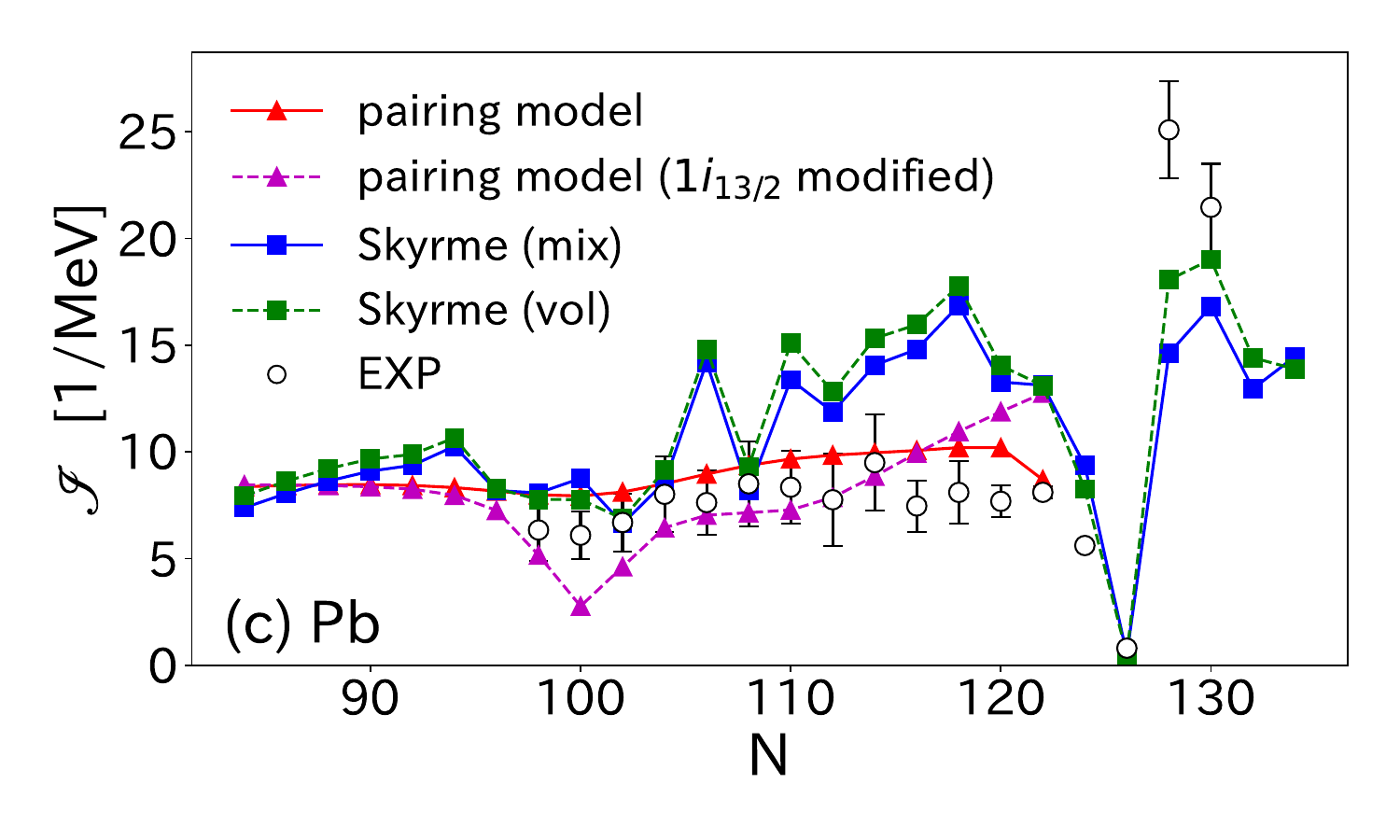}
\end{center}
\caption{Calculated P-MoI ${\mathscr I}$ for (a) Ni, (b) Sn, and (c) Pb isotopes, with the pairing Hamiltonian (red triangles), the Skyrme EDF (blue squares for mixed-type and green for volume-type pairing EDF), and the experimental values \cite{Wang2021} (black circles).
In the panel (c), those with a modified single-particle energy of $1i_{13/2}$ are shown (magenta triangles). See text for details.
\label{fig:NMOI}}
\end{figure}

We calculate the P-MoI using Eq.~(\ref{eq:pairingMOIapr}) from the ground-state energies of 
the isotopes with $N-2, N$, and $N+2$ neutrons.
The results for Ni, Sn, and Pb isotopes are shown in Fig. \ref{fig:NMOI}.
The experimental values are evaluated using the binding energies taken from AME2020 \cite{Wang2021}.
As a general trend, the P-MoI increases with the neutron number
for Ni and Sn isotopes except for those with the magic numbers.
The origin of this tendency will be discussed in Sec.~\ref{sec:number-dependence}.

This increasing tendency with the neutron number
is weak in the Pb isotopes.
The main difference in the single-particle-level structure between Pb and the other two (Ni and Sn) is the position of the intruder orbits:
The $1g_{9/2}$ and $1h_{11/2}$ orbits in Ni and Sn isotopes are located 
near the end of the shell with the magic numbers of
$N=50$ and $82$, respectively,
whereas the $1i_{13/2}$ in Pb is located in the middle of the shell
in the model space as shown in Table~\ref{table:singleeng}.
Since the high-$j$ intruder orbits have large degeneracy,
their position may significantly affect the pairing property.
To see this effect, we change the energy of the $1i_{13/2}$ intruder orbit of the Pb isotopes from $-9.335$ MeV to the last orbit in the shell, $-7.000$ MeV.
The result is shown with magenta triangles in Fig.~\ref{fig:NMOI}(c).
The P-MoI with the modified $1i_{13/2}$ energy has an apparent increasing trend toward the shell gap at $N=126$.
The high-$j$ intruder orbits are responsible for the increasing behavior of the P-MoI as a function of neutron number.
Since the experimental values of the P-MoI in Pb isotopes
are almost constant for different neutron numbers in $N<126$,
the position of the intruder $1i_{13/2}$ orbit in Table~\ref{table:singleeng} is consistent with the experimental data,
which suggests its location around the middle of the shell between $N=82$ and 126.

The Skyrme-DFT calculations with the SLy4 parameters produce
consistent results with those of the pairing Hamiltonian.
Small differences in the Sn isotopes may be due to differences in
the order of the single-particle energies between the Skyrme DFT and
those in Table~\ref{table:singleeng}.
In Fig.~\ref{fig:NMOI}(c),
we observe a staggering behavior in
the Skyrme-DFT results for Pb isotopes.
The same kind of behaviors have been also observed in a former study
\cite{Hinohara2018}.
This may be due to a small error involved in the calculated ground-state energy.
The total binding energy is approximately proportional to the mass number
$A$, producing about $B\approx 1.6$ GeV for $A\approx 200$,
while the two-nucleon shell gap $\delta_{2n}$ of Eq.~(\ref{eq:DS2n})
is order of 10 keV.
Therefore, 
a small error of $10^{-3}$ \% in the calculation of the total binding energy
could significantly impact $\delta_{2n}$.
The double binding-energy difference of
Eq.~(\ref{eq:pairingMOIapr}) for heavy nuclei
requires high precision in the calculation of the ground-state energy.
Reference~\cite{Hinohara2018} shows that
in contrast to the double binding-energy difference,
the Thouless-Valatin P-MoI calculated with the quasiparticle-random-phase approximation indicates a smooth behavior as a function of the neutron number.

The P-MoI of these proton-magic spherical nuclei
indicate strong dip behaviors at the neutron magic numbers
in Fig.~\ref{fig:NMOI}.
This behavior has been known in terms of the empirical shell gap
of the two-neutron separation energy $\delta_{2n}$.
In contrast,
for the superfluid nuclei of open-shell configurations,
$\delta_{2n}$ is directly connected to the P-MoI as
${\mathscr I}(N)=4/\delta_{2n}(N)$,
and should be explained in terms of the pairing properties of nuclei.

\subsection{Particle-number dependence of P-MoI}
\label{sec:number-dependence}

\begin{figure}[H]
\begin{center}
\includegraphics[width=80mm]{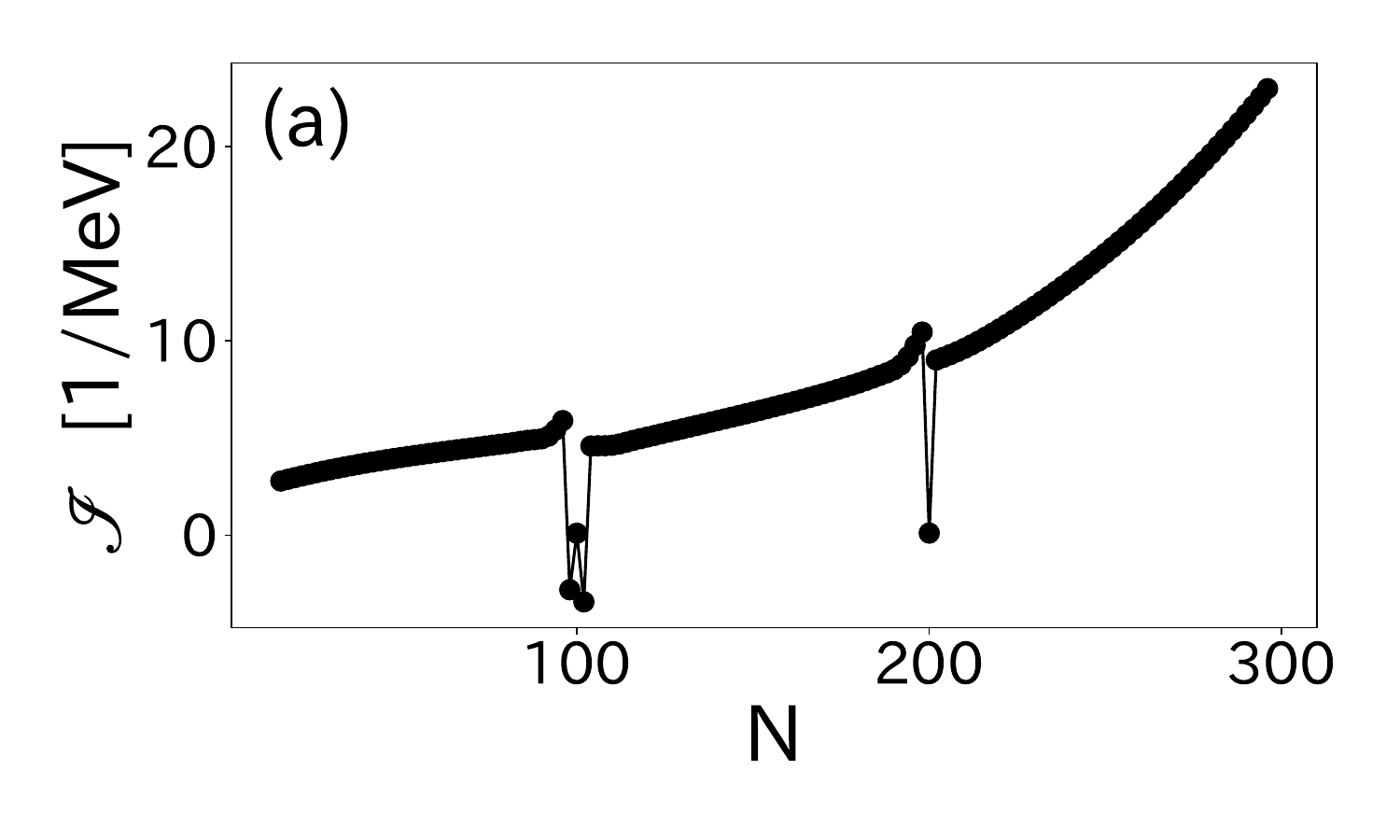} \\
\includegraphics[width=80mm]{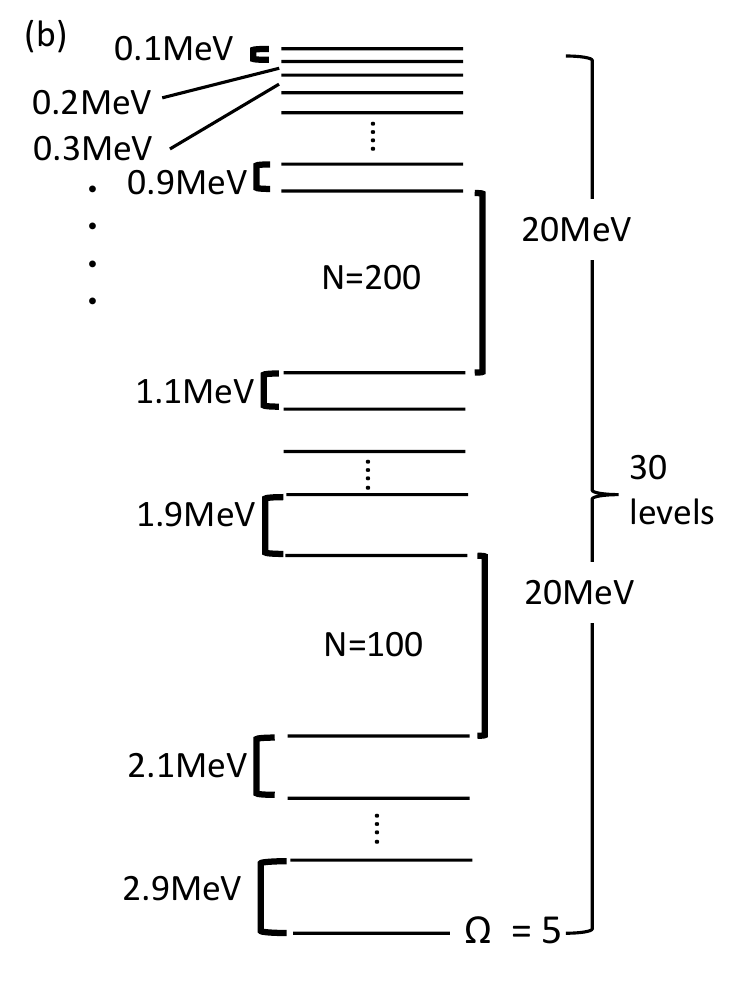}
\end{center}
\caption{(a) The P-MoI as a function of the neutron number for the pair model given with the single-particle levels in the case (4) in the text. (b) Single-particle energies of the case (4). See text for details.
}
\label{fig:NMOI30}
\end{figure}

To understand the trend of the P-MoI, we adopt a larger model space with 30 single-particle levels for the pairing model of Eq. (\ref{H}).
All the 30 levels have 10-fold degeneracy ($\Omega_j=5$).
This provides an ideal environment for studying the effects of the shell structure on the pairing properties.
We study the following four cases;
\begin{enumerate}
\item[(1)]
Equal spacing of the single-particle energy intervals,
$\Delta e = 1$ MeV,
and the constant force strength, $G=0.3$ MeV.
\item[(2)] The constant $G$, but smaller energy spacing for higher levels,
$\Delta e=2.9,\ 2.8,\ \cdots,{\rm and}~  0.1$ MeV,
from the bottom to the top.
\item[(3)] Equal spacing $\Delta e = 1$ MeV,
but the force strength, $G=G_0/N$ with $G_0=0.3$ MeV.
\item[(4)] Smaller energy spacing for higher levels,
$\Delta e=2.9,\ 2.8,\ \cdots,{\rm and}~0.1$ MeV.
In addition, 20 MeV shell gaps
inserted between the 10th and 11th levels
and between the 20th and 21st levels. See Fig.~\ref{fig:NMOI30} (b).
The force strength, $G=G_0/N$ with $G_0=0.3$ MeV.
\end{enumerate}

In the case (1),
the P-MoI is approximately constant for all the neutron numbers
except for the edge of the model space ($N\approx 0$ and $N\approx 300$).
In other words, the difference of the two-neutron separation energies,
$\delta_{2n}(N)$, does not depend on $N$.
In the case (2),
The P-MoI shows an upward trend with the neutron number:
higher level densities produce larger P-MoI.
In the case (3),
the P-MoI increases with the neutron number:
Smaller $G$ produces larger P-MoI.
The results in the case (4) is shown in Fig.~\ref{fig:NMOI30}(a).
The singular behavior is seen at the neutron numbers corresponding to the magic number ($N=100$ and 200).
The dips at the magic numbers also exist in Ni, Sn, and Pb isotopes
(Fig.~\ref{fig:NMOI}).

We conclude that both the neutron-number dependence $1/N$ of the pairing force strength and the increase of the level density are responsible for the overall upward trend of the P-MoI.
At the neutron numbers corresponding to the shell gaps,
there are dips in the P-MoI.

The neutron empirical shell gaps $\delta_{2n}(Z)$
as a function of the proton number ($Z$)
for isotones with the neutron magic numbers have a peak
at the doubly magic nuclei \cite{Lunney2003}.
The neutron empirical shell gap is a useful tool to confirm
doubly magic nuclei, for example, $^{78}$Ni \cite{Manea2023, Welker2017}.
In the present study, we find that the P-MoI is affected by the underlying single-particle level density and the pairing force strength.
Therefore, the P-MoI in the open-shell superfluid nuclei may provide useful information on the shell structure and the pairing properties.

\subsection{Order-parameter dependence of P-MoI\label{sec:PRMOI-orderparameter}}

\begin{figure}[tbp]
\begin{center}
\includegraphics[width=80mm]{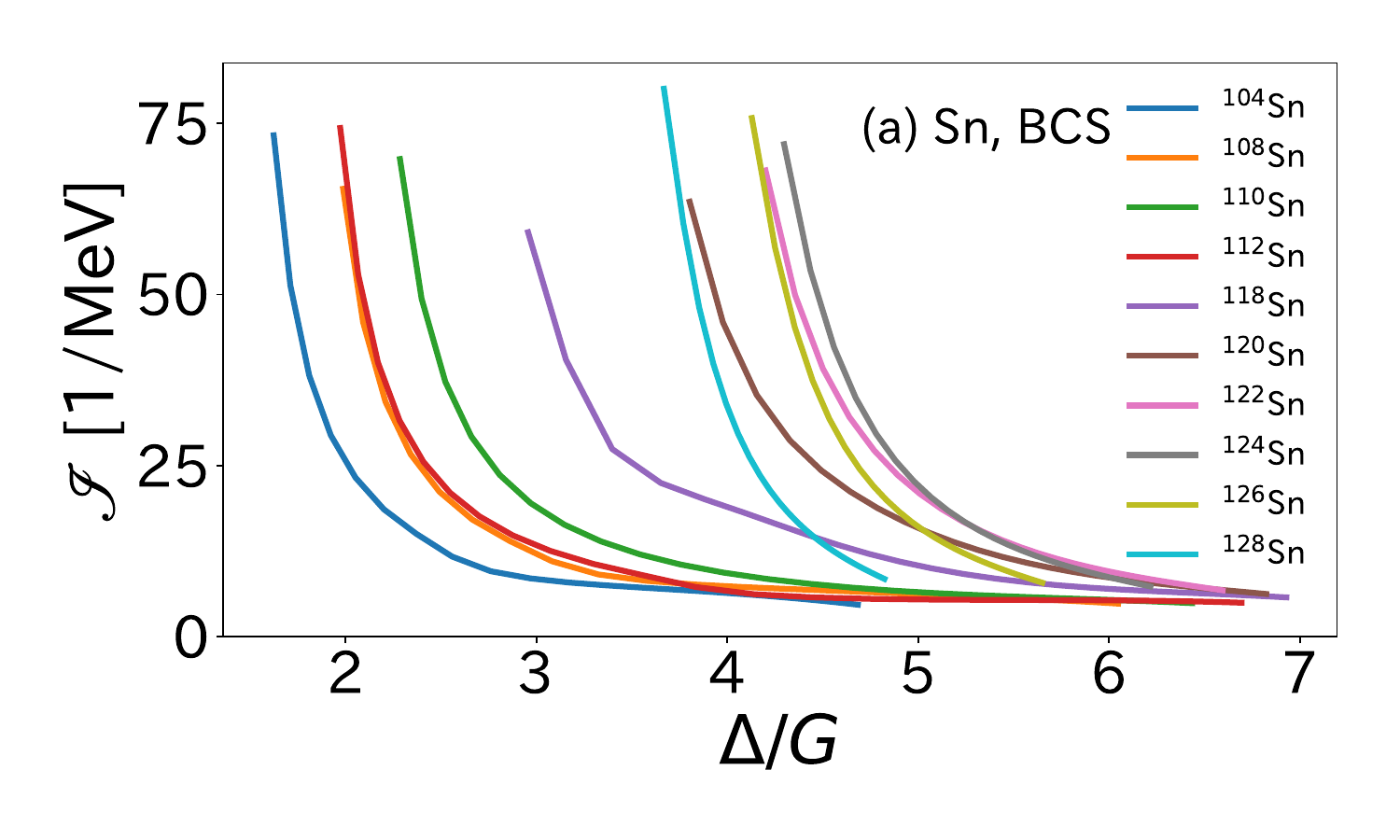} \\
\includegraphics[width=80mm]{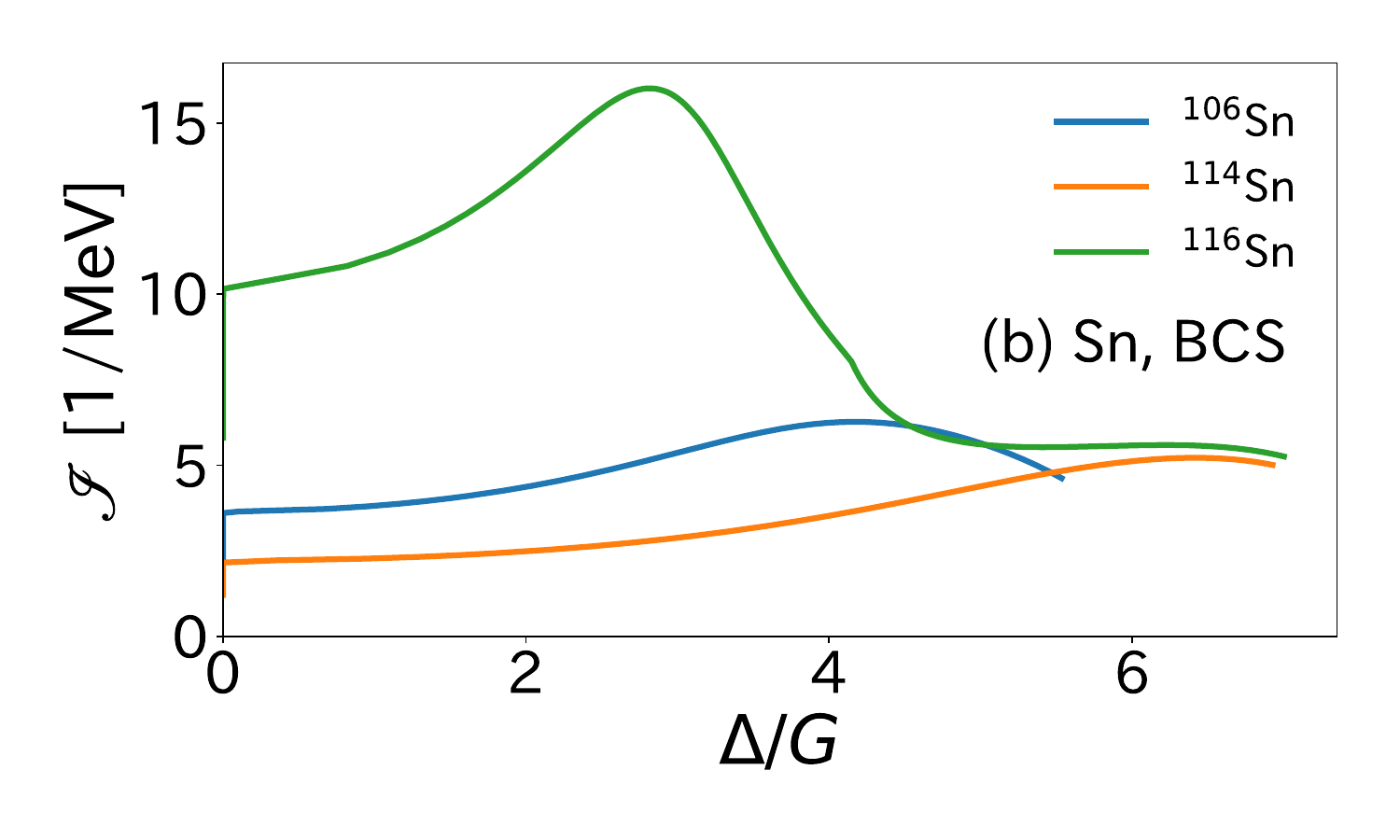}
\end{center}
\caption{The P-MoI as a function of the pair amplitude for Sn isotopes,
calculated with the BCS approximation using the pairing Hamiltonian.
(a) open-shell nuclei and (b) closed-subshell nuclei.
\label{fig:DMOISNBCS}}
\end{figure}

Next, we study the dependence of the P-MoI on the pair amplitude, the order parameter for the normal to the superconducting phase transition.
In the BCS calculation of the pair model, the pair amplitude is
given by $\Psi=\langle\hat{A}\rangle=\Delta/G$.
Figure~\ref{fig:DMOISNBCS} shows the P-MoI as a function of $\Delta/G$,
calculated with different values of the pairing force strength $G$ in the pairing Hamiltonian.
In open-shell Sn isotopes, we clearly see a negative correlation between the P-MoI and the pair amplitude.
In contrast, in the closed-subshell Sn isotopes ($2d_{5/2}$-closed $^{106}$Sn, $1g_{7/2}$-closed $^{114}$Sn, and $3s_{1/2}$-closed $^{116}$Sn), 
when the force strength is smaller than the critical value $G<G_c$, the P-MoI calculated with the double binding-energy difference increases because the energy of the neighboring $N\pm 2$ nuclei changes.
Note that for $^{114}$Sn and $^{116}$Sn, one of the neighboring even-$N$ nuclei is also subshell-closed.
For $G>G_c$, the P-MoI increases in a region up to a certain value of $\Delta/G$, then starts decreasing beyond that value.
We obtain similar results for Ni isotopes.

\begin{figure}[tbh]
\begin{tabular}{cc}
\includegraphics[width=80mm]{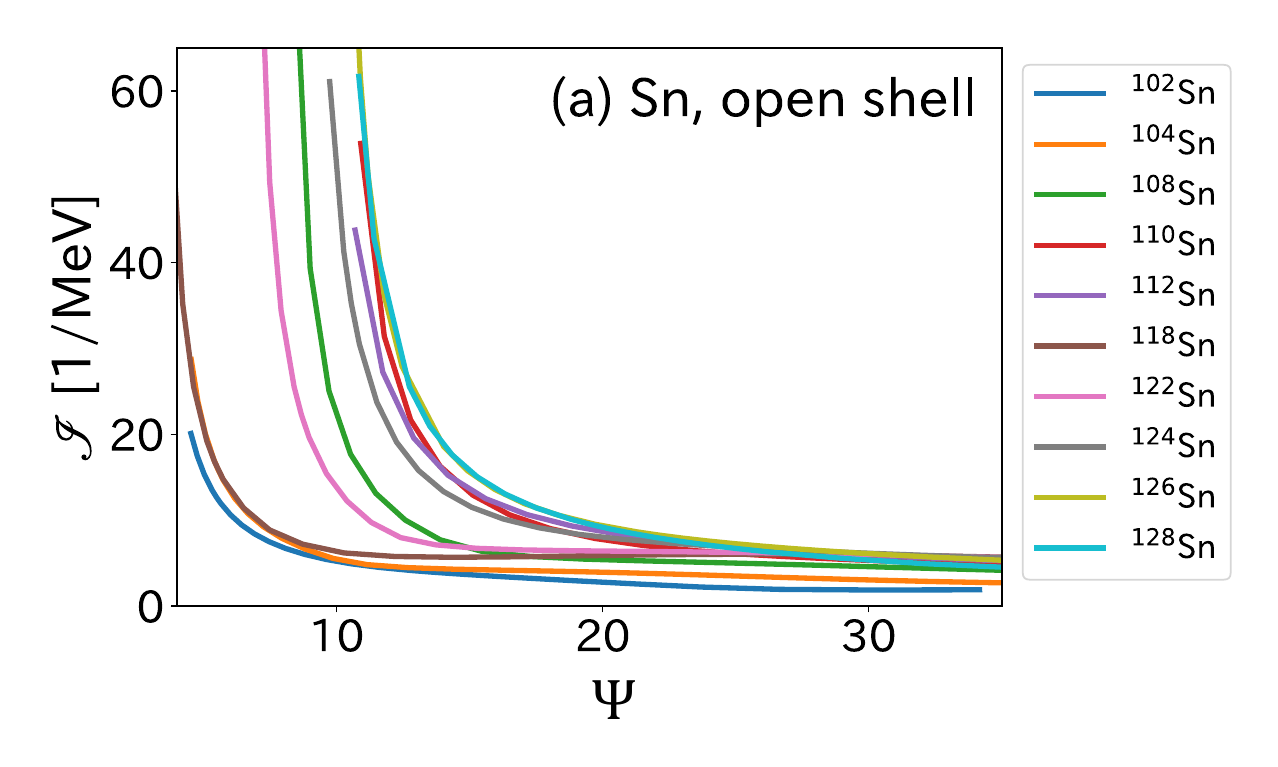} \\
\vspace{-10pt}
\includegraphics[width=75mm]{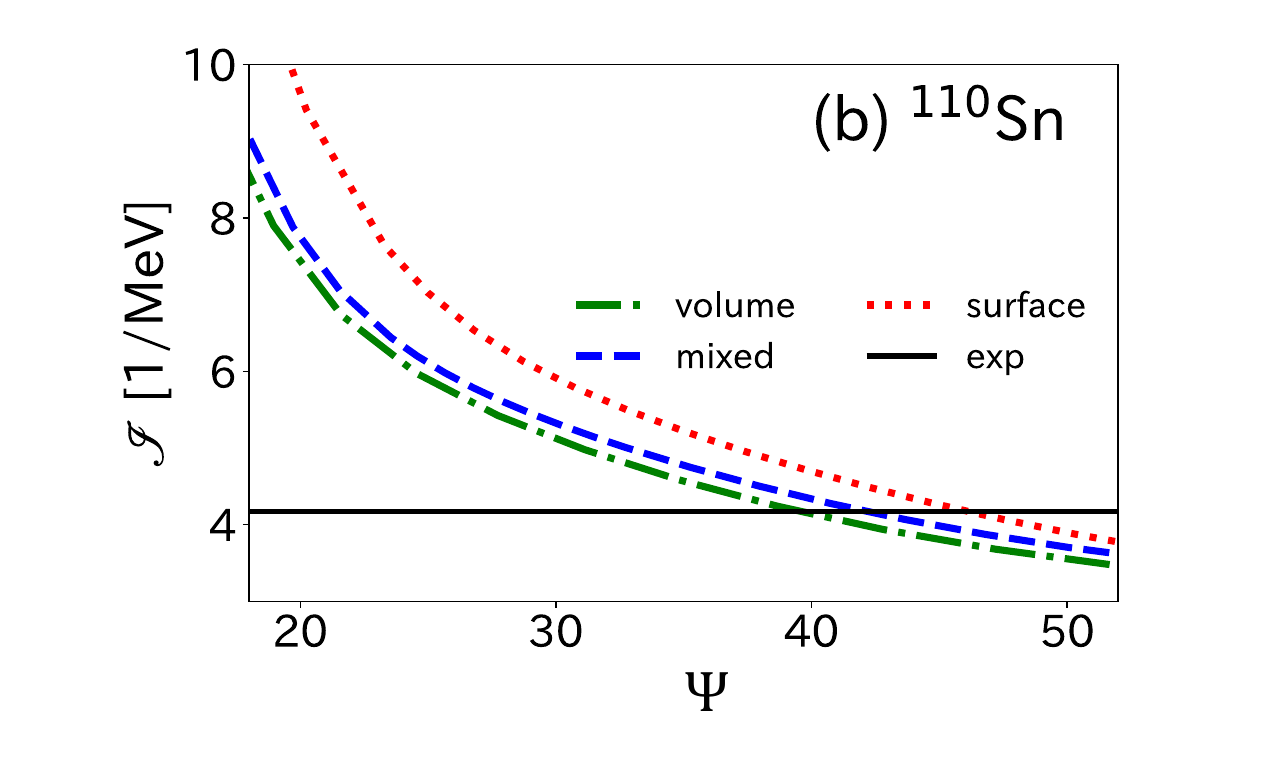} \\
\vspace{-10pt}
\includegraphics[width=66mm]{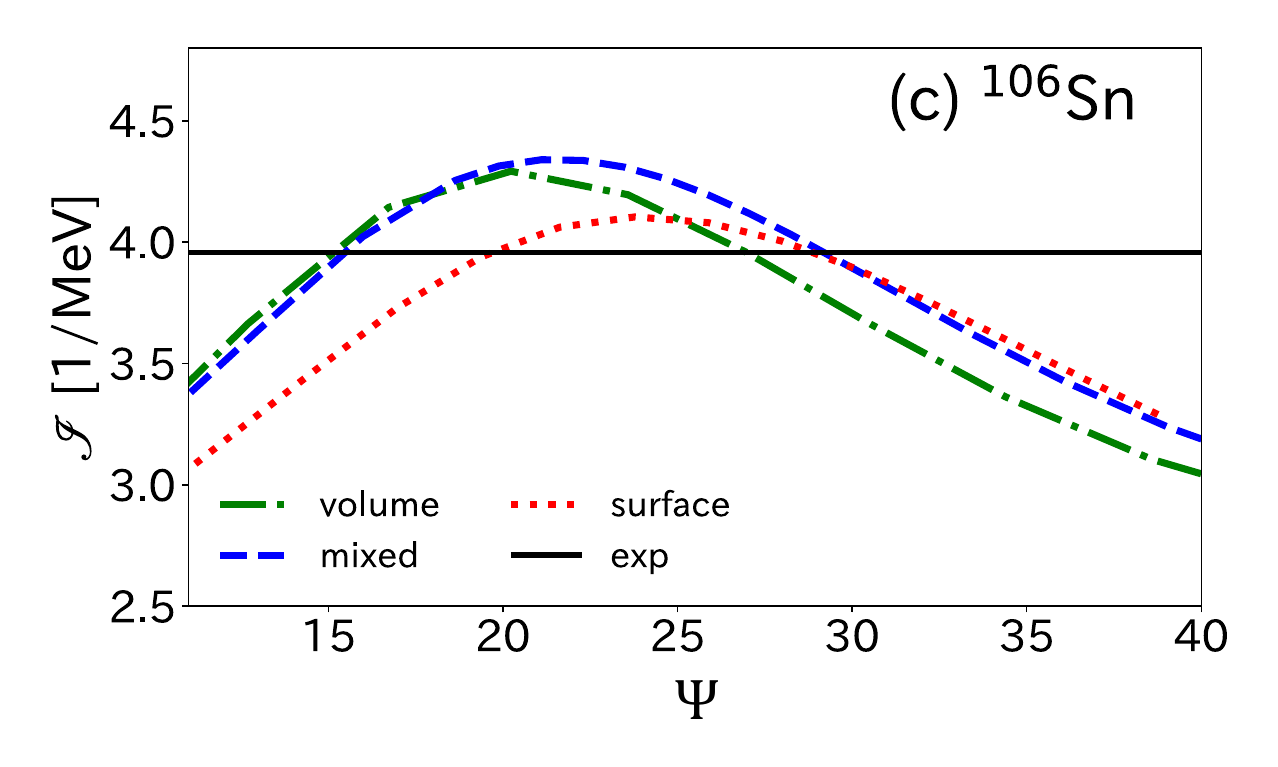} \\
\vspace{-10pt}
\includegraphics[width=75mm]{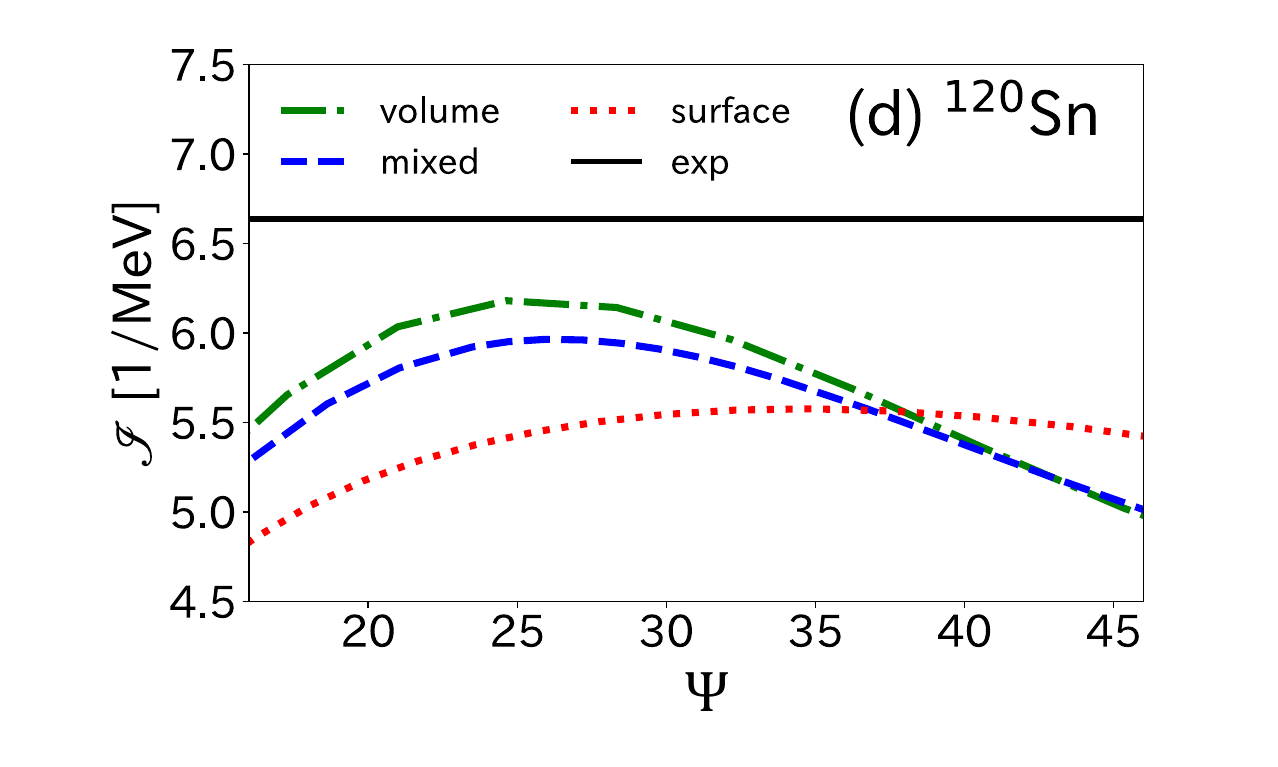} \\
\end{tabular}
\vspace{-10pt}
\caption{
Calculated P-MoI as a function of the pair amplitudes,
using the Skyrme EDF and the pairing EDF with different coupling constant $V_0$.
(a) P-MoI for all the open-shell Sn isotopes in $50\le N \le 82$ using the mixed-type pairing EDF.
(b) P-MoI compared with the experimental value (horizontal solid line) for 
$^{110}$Sn.
(c) The same as (b) but for $^{106}$Sn.
(d) The same as (b) but for $^{120}$Sn.
\label{fig:DMOISNSkyrme}
}
\end{figure}

We conduct the same analysis using the Skyrme EDF with the volume-type and mixed-type density-dependent pairing EDFs.
In the case of the local pairing EDFs, the quasiparticle cutoff may cause a sudden change of the model space when changing the pairing strength, which can result in discontinuous behavior in the HFB energy and the P-MoI.
Thus in this analysis, we set the common model space without the quasiparticle cutoff and 
use a relatively small single-particle model space with
14 major shells.
The number of the quasiparticle states within the model space
for $^{116}$Sn
is 680,
which is close to 687, the number of those
with the model space of 20 major shells and the 60 MeV cutoff for the quasiparticle states.
The neutron pair amplitude in the case of the Skyrme EDF is defined by integrating the local pair density 
\begin{align}
\Psi =  \int d\textbf{r} \tilde{\rho}_n(\textbf{r}).
\end{align}
The correlation between the P-MoI and the pair amplitude $\Psi$
is shown in Fig.~\ref{fig:DMOISNSkyrme}.
The basic properties are
the same as the case of the pairing Hamiltonian;
in the open-shell Sn isotopes, there is a negative correlation 
[Fig.~\ref{fig:DMOISNSkyrme}(a)], whereas 
in the closed-subshell Sn isotopes
($^{106,114,116,120}$Sn in the case of SLy4 EDF),
a positive correlation for small values of $\Psi$
and a subsequent negative correlation for larger values of $\Psi$.

Figure~\ref{fig:DMOISNSkyrme}(b) shows 
that the experimental value of the P-MoI
in the open-shell nucleus $^{110}$Sn can constrain the value of
the pair amplitude within about 10\% difference 
depending on different density dependences of the pairing EDF.
On the other hand, in the closed-subshell nuclei, for example, $^{106}$Sn, there are two possible values for each pairing type (Fig.~\ref{fig:DMOISNSkyrme}(c)).
$^{120}$Sn is also subshell closed in the Skyrme-DFT calculation
with the SLy4 parameters.
In $^{120}$Sn shown in Fig.~\ref{fig:DMOISNSkyrme}(d),
the present model cannot reproduce
the experimental P-MoI with any type of pairing EDF.
The neutron OES of $^{120}$Sn is often used to fit the coupling constant of the pairing EDF.
It is always possible to fit the pairing gap with the experimental OES.
However, the present analysis shows that it may have difficulty to reproduce 
the P-MoI for nuclei with closed-subshell configurations.
It may be more appropriate to choose an open-shell nucleus
to determine the coupling constant.

The negative correlation between the order parameter and the P-MoI
in open-shell nuclei is clearly different from
the positive correlation between the deformation and the MoI
for the spatial rotation.
The spatial rotation and the pair rotation have many properties in common
as they have the common origin as the emergent degrees of freedom
associated with the spontaneous symmetry breaking.
Nevertheless, the relation between the MoI and the order parameter
does not coincide.

\subsection{Analysis based on the BCS approximation}

The negative correlation between the pair amplitude and the P-MoI
in open-shell nuclei can be qualitatively explained
in terms of the BCS approximation.
A preliminary version of the discussion has been reported in Ref.~\cite{EPJWebConf.306.01006}.
Introducing the single-particle level density $\rho(\epsilon)$,
the particle number $N$ is given as
\begin{align}
N(\lambda,\Delta)&=\int_{-\infty}^\infty
\rho(\epsilon) V^2(\epsilon-\lambda,\Delta)d\epsilon,\\
V^2(\epsilon-\lambda,\Delta)&=\frac{1}{2}
\left( 1-\frac{\epsilon-\lambda}{\sqrt{(\epsilon-\lambda)^2+\Delta^2}} \right), \label{eq:occupation}
\end{align}
where $V^2(\epsilon-\lambda,\Delta)$ is the occupation probability of the BCS formulation.
We do not solve the gap equation in the present analysis.
It is enough to assume that given $(N,\Delta)$
can be achieved by tuning $\lambda$ and the pairing force strength $G$.
Using the relation $\lambda=dE/dN$,
the P-MoI is given by 
\begin{align}
\mathscr{I}(\lambda,\Delta)&=\left(\frac{d^2E}{dN^2}\right)^{-1}
=\frac{dN}{d\lambda}\nonumber \\
&=\int_{-\infty}^{\infty}
\rho(\epsilon) \frac{\partial V^2(\epsilon-\lambda,\Delta)}{\partial\lambda} 
d\epsilon\nonumber\\
&=\int_{-\infty}^{\infty}
\rho(\epsilon) f(\epsilon-\lambda,\Delta)
d\epsilon,
\label{MOIpair} \\
f(\epsilon-\lambda,\Delta) &\equiv
\frac{1}{2}
\frac{\Delta^{2}}{[(\epsilon-\lambda)^{2}+\Delta^{2}]^{3/2}}. 
\label{eq:f}
\end{align}
Here, we neglect a contribution from the $\lambda$-dependence of $\Delta$.
The integrand of Eq.~(\ref{MOIpair}) is a product of 
the level density $\rho(\epsilon)$
and $f(\epsilon-\lambda,\Delta)$.
The function $f(\epsilon-\lambda,\Delta)$
is strongly peaked at $\epsilon=\lambda$ and
decreases with the pairing gap
in the energy region near the Fermi energy, $\epsilon\approx\lambda$.
The expression of Eq. (\ref{MOIpair})
may explain the increasing behavior of the P-MoI
with the neutron number for open-shell nuclei in Fig.~\ref{fig:NMOI}.
Suppose that the level density monotonically increases with the single-particle energy.
At the beginning of the shell, the pairing gap $\Delta$ increases with the neutron number,
which leads to a decrease in  $f(0,\Delta)=1/(2\Delta)$.
The competing effects between $\rho(\lambda)$ and $f(0,\Delta)$
keeps the product $\rho(\lambda)f(0,\Delta)$ roughly constant,
thus, the P-MoI does not change much with respect to the neutron number.
In the latter part of the shell beyond the midpoint
where the gap decreases with the neutron number,
both $\rho(\lambda)$ and $f(0,\Delta)$ coherently contribute to
the increase in the P-MoI.
Equation~(\ref{MOIpair}) also explains why there is a strong dip
near the magic number, observed in Fig.~\ref{fig:NMOI}.
It is simply because $\rho(\lambda)\approx 0$ at the magic number.

We can conclude that the P-MoI is affected by
both the level density and the pairing gap (pair amplitude).
Equation~(\ref{MOIpair}) provides a useful insight into the property of
the P-MoI:
When the level density drastically changes near the Fermi energy $\lambda$,
it reflects the single-particle shell gap.
On the other hand, if the level density is a smooth function of $\epsilon$
at $\epsilon\approx\lambda$,
the pair rotation, the origin of which is the gauge symmetry breaking,
is a dominant content in P-MoI.

\subsection{Orbital decomposition of the P-MoI}

In bound finite nuclei, the level density near the Fermi energy is discrete.
In this case, the level density and the particle number should be
replaced by
\begin{align}
 \rho(\epsilon) &= 2 \sum_j \Omega_j \delta( \epsilon - \epsilon_j) , \\
N(\lambda,\Delta)& = 
2 \sum_j \Omega_j V^2(\epsilon_j-\lambda,\Delta) ,
\end{align}
where $2\Omega_j$ represents the degeneracy of
the single-particle level $j$.
The P-MoI is thus given by 
\begin{align}
\mathscr{I}(\lambda,\Delta)&=\frac{dN}{d\lambda}
= 2\sum_j \Omega_j \frac{\partial V^2(\epsilon_j-\lambda,\Delta)}{\partial\lambda}  \nonumber \\
&=\frac{1}{\Delta} \sum_j \Omega_j
\biggr[ 1 + \biggr( \frac{\epsilon_j-\lambda}{\Delta}\biggr)^2\biggr]^{-\frac{3}{2}}.
\label{eq:MOIpair}
\end{align}
We should note that we neglect
the self-consistent relation between $\lambda$ and $\Delta$.
$\lambda$ should depend on $\Delta$ and vice versa,
however, this is not taken into account in the present discussion.
Therefore, the detailed properties obtained with 
Eq.~(\ref{eq:MOIpair})
may not be true for realistic cases.
Nevertheless, the following discussion qualitatively holds and
is useful for the analysis of the pairing rotation.

In order to understand the property of Eq. (\ref{eq:MOIpair}),
we consider three extreme cases.
\newline\noindent
\underline{Large $\Delta$ limit}:
$\Delta\gg \displaystyle\max_j|\epsilon_j-\lambda|$ leads to
\begin{align}
 \mathscr{I}(\lambda,\Delta) \approx \frac{1}{\Delta}
 \sum_j \Omega_j \biggr[
  1 - \frac{3}{2} \biggr( \frac{\epsilon_j-\lambda}{\Delta}\biggr)^2
 \biggr]
 \approx \frac{1}{\Delta}\sum_j\Omega_j .
\end{align}
The P-MoI is proportional to $\Delta^{-1}$.
This is consistent with those of the open-shell nuclei
in Figs.~\ref{fig:DMOISNBCS} and \ref{fig:DMOISNSkyrme}.
\newline\noindent
\underline{Small $\Delta$ limit in closed-(sub)shell nuclei}:
$\Delta \ll \displaystyle\min_j|\epsilon_j-\lambda|$ leads to
\begin{align}
 \mathscr{I}(\lambda,\Delta) \approx \Delta^2 \sum_j \frac{\Omega_j}{|\epsilon_j - \lambda|^3}. \label{eq:Delta0closedshell}
\end{align}
The P-MoI is proportional to $\Delta^2$.
This condition can be satisfied for closed-(sub)shell nuclei
with the small $\Delta$ limit.
However, in Fig.~\ref{fig:DMOISNBCS},
the dependence $\mathscr{I}\propto\Delta^2$ is not realized.
This is due to the fact that
the double binding-energy difference of $\Delta N=\pm 2$
is not consistent with the present analysis.
See Sec.~\ref{sec:DBED}.
In a realistic situation,
the shell gap provides a nonzero P-MoI value at $\Delta\rightarrow 0$.
Nevertheless, at small values of $\Delta$,
we may approximate them by parabolic curves.
\newline\noindent
\underline{Small $\Delta$ limit in open-shell nuclei}:
In the small $\Delta$ limit in open-shell nuclei,
$\lambda$ approaches the Fermi-level energy $\epsilon_{\rm F}$.
The pairing correlation is active only within the Fermi level.
The P-MoI is thus given by
\begin{align}
 \mathscr{I}(\lambda,\Delta) \approx 
 \frac{\Omega_{\rm F}}{\Delta}\left[
 4 V_{\rm F}^2 (1 - V_{\rm F}^2)
 \right]^{\frac{3}{2}} ,
 \label{eq:Deltato0openshell}
\end{align}
where $2\Omega_{\rm F}$ is the degeneracy of the Fermi level, and 
its occupation probability
\begin{align}
 V_{\rm F}^2 = \frac{1}{2} \biggr(
 1 - \frac{\epsilon_{\rm F} - \lambda}{\sqrt{ (\epsilon_{\rm F} - \lambda)^2 + \Delta^2}}
 \biggr)
 \approx \frac{N_{\rm F}}{2\Omega_{\rm F}},
\end{align}
where $N_{\rm F}$ is the number of particles in the Fermi level.
Note that the formula of Eq.~(\ref{eq:Deltato0openshell})
is different from the exact expression
for the single-level case\footnote{The discrepancy is due to lack of the self-consistency between $\lambda$ and $\Delta$.},
though the basic properties,
such as $\mathscr{I}\approx G^{-1}$,
are reproduced.

The derivative of the P-MoI with respect 
to the pairing gap may be useful for understanding the behavior of the P-MoI against the pairing gap
\begin{align}
 \frac{\partial\mathscr{I}(\lambda,\Delta)}{\partial\Delta}
  = 
  - \frac{1}{\Delta^2} \sum_j
  \frac{\displaystyle 1 - 2\biggr(\frac{\epsilon_j-\lambda}{\Delta}\biggr)^2}
  {\displaystyle \biggr[1 + \biggr(\frac{\epsilon_j-\lambda}{\Delta}\biggr)^2\biggr]^{5/2}}. \label{eq:MOIpair-derivative}
\end{align}
This expression shows how each single-particle state
contributes to the derivative of P-MoI.
The sign of the contribution to the derivative is determined by the numerator; the levels close to the Fermi energy satisfying $|\epsilon_j-\lambda|<\Delta/\sqrt{2}$
give negative contributions to the derivative, while the levels off 
the Fermi energy $|\epsilon_j-\lambda|>\Delta/\sqrt{2}$
have a positive contribution to the derivative.
By increasing $\Delta$, the number of the single-particle levels that have negative contributions to the derivative increases.
Therefore, in the large $\Delta$ limit, there is always a negative
correlation between the P-MoI and the pairing gap.

The P-MoI of Eq.~(\ref{eq:MOIpair})
coincides with the Belyaev cranking inertia for the pair
rotation in the case of the pairing Hamiltonian
\cite{Inglis1956,Belyaev1965}
\begin{align}
\mathscr{I}_{\rm Belyaev}=
4\sum_{j}\frac{\Omega_j U_j^{2}V_j^{2}}{E_j}.
\end{align}
Here 
$E_{j}$ is the quasiparticle energy.
By inserting Eq.~(\ref{eq:occupation}) into $V_j^2$ and $U_j^2 = 1 - V_j^2$
in the above equation, Eq.~(\ref{eq:MOIpair}) is obtained.
The equivalence with Belyaev's inertia suggests that the residual correlations may be missing in the present analysis.
In the case of the proton pairing rotation,
the residual effect is as large as a factor of 2-3
due to the Coulomb EDF~\cite{Hinohara2016}.

\begin{figure}[tbh]
\begin{center}
\includegraphics[width=84mm]{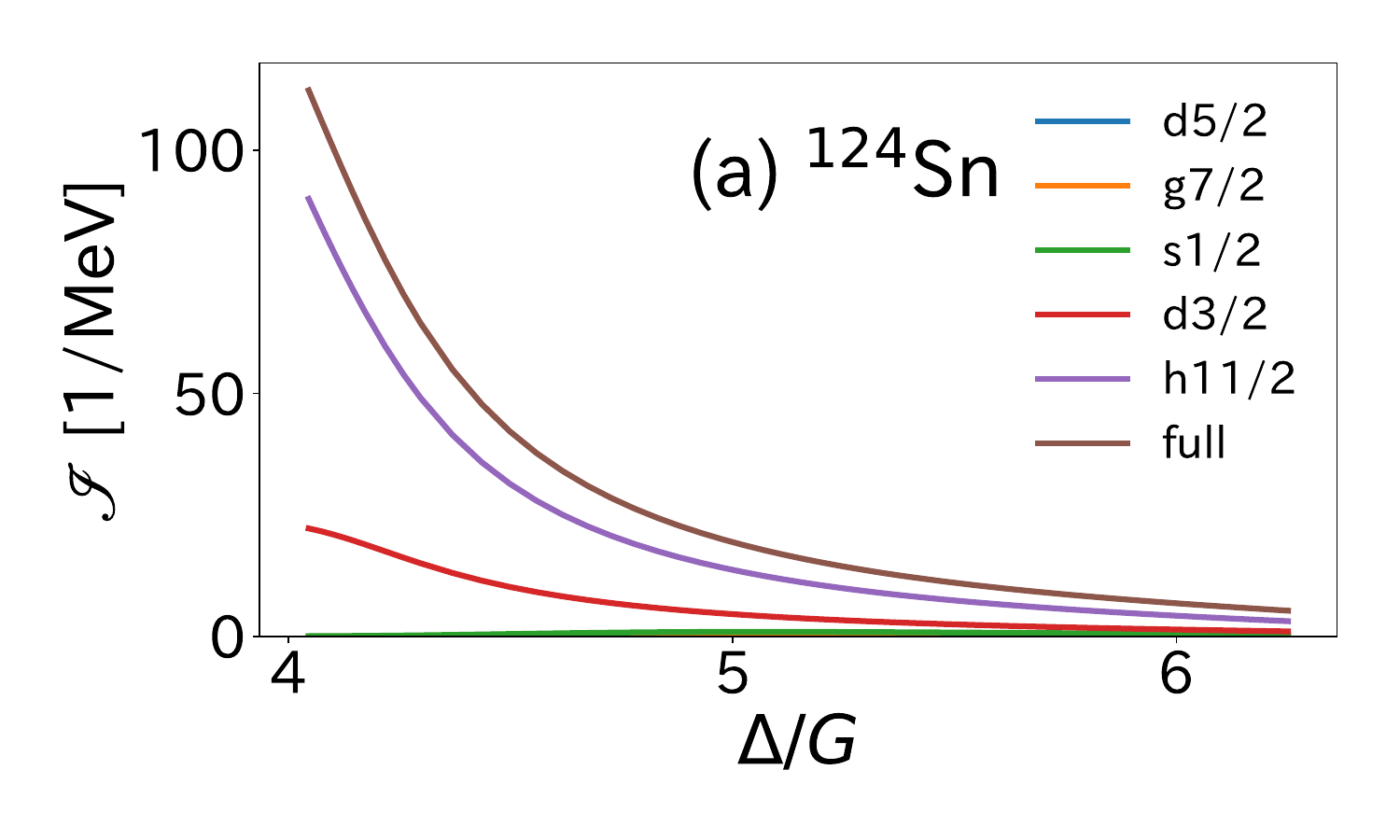}\\
\includegraphics[width=80mm]{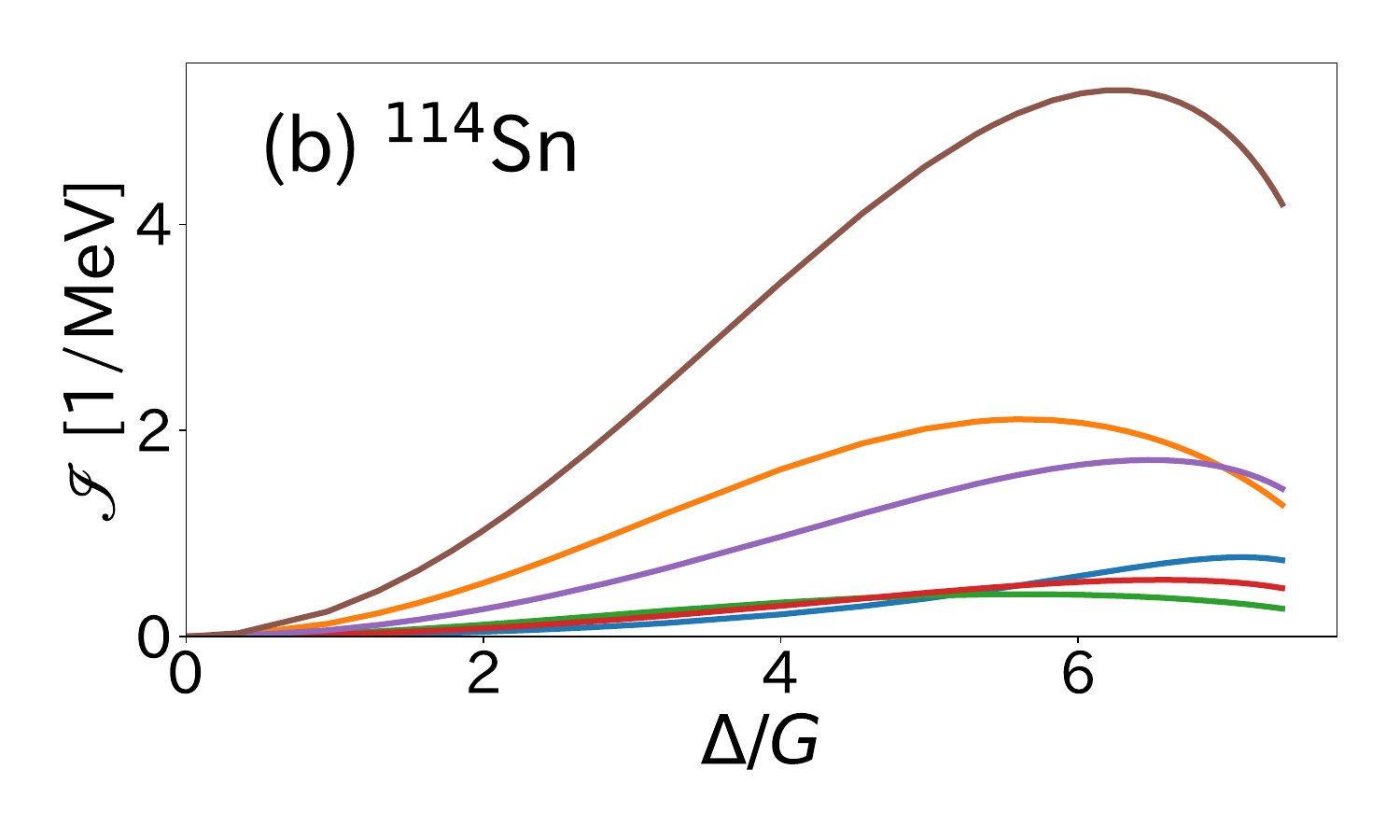}
\end{center}
\caption{The P-MoI and its orbital decomposition
as a function of the pair amplitude,
calculated with the pairing Hamiltonian.
(a) Open-shell nucleus $^{124}$Sn.
(b) Closed-subshell nucleus $^{114}$Sn. 
}
\label{fig:MOI-delta-splevel}
\end{figure}

Figure~\ref{fig:MOI-delta-splevel} shows the orbital decomposition of the P-MoI
for the open-shell $^{124}$Sn and the closed-subshell $^{114}$Sn nuclei
in the case of the pairing Hamiltonian.
In $^{124}$Sn, the largest contribution
comes from $1h_{11/2}$ level that is close to the Fermi energy.
As we have seen in Eq.~(\ref{eq:Deltato0openshell}), the contribution from $1h_{11/2}$ is dominant and becomes exclusive in the limit of $\Delta\rightarrow 0$.
The relative importance of each single-particle level can be understood in terms of the degeneracy $\Omega_j$ and
the dimensionless quantity $(\epsilon_j-\lambda)/\Delta$.
In $^{114}$Sn, the main contribution to the P-MoI comes from the $1g_{7/2}$ orbit that is the closest to the Fermi energy.
In the limit of the strong pair amplitude,
we see that $1h_{11/2}$ with the highest degeneracy
has the largest contribution. 

\begin{figure}[tbh]
\begin{center}
\includegraphics[width=85mm]{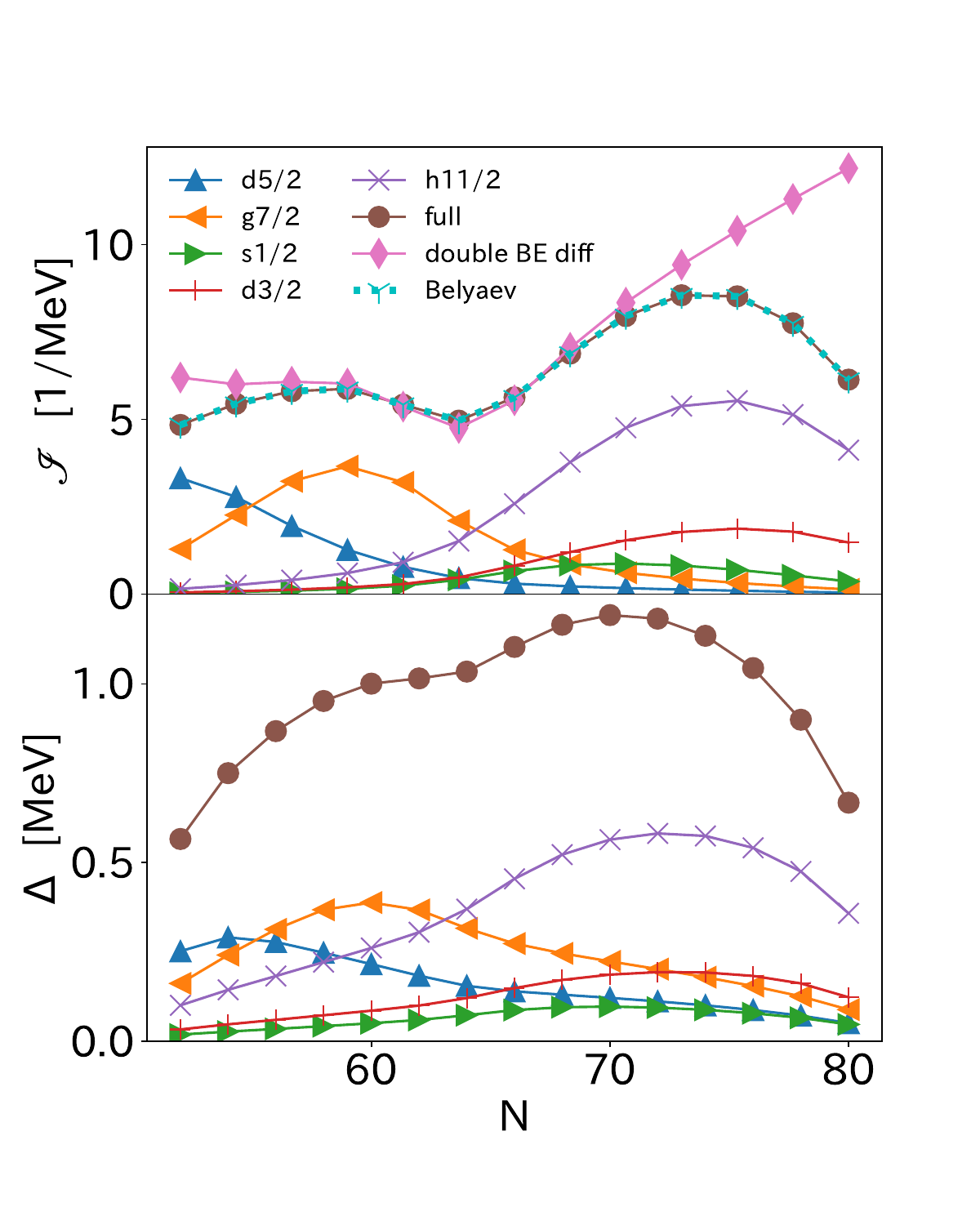}
\end{center}
\caption{(Upper panel)
The P-MoI and its orbital decomposition as a function
of the neutron number for Sn isotopes,
calculated from the double binding-energy differences (diamonds)
and the Belyaev approximation (dotted line)
using the pairing Hamiltonian.
(Lower) Pairing gap and its orbital decomposition. \label{fig:MOI-N} } 
\end{figure}

Figure~\ref{fig:MOI-N} shows the 
orbital decomposition of the P-MoI and the pairing gap.
Although they have opposite correlations against the order parameter $\Delta/G$,
their composition in terms of
the single-particle levels are 
very similar.
In the lighter region, $d_{5/2}$ and $g_{7/2}$ orbits 
have a major contribution,
whereas those of $h_{11/2}$ and $d_{3/2}$ orbits become dominant in the heavier region.
The relatively flat behavior of the P-MoI in $54\le N \le 64$ can be understood in terms of the splitting of the peaks of $d_{5/2}$ (peaked at $N=54$) 
and $g_{7/2}$ (peaked at $N=60$) contributions.
The increasing behavior of the P-MoI in $64\le N\le 80$
is due to the same location of the peaks of $h_{11/2}$ and $d_{3/2}$ at around $N=74$,
that is due to the almost degenerated single-particle energies of $h_{11/2}$ and $d_{3/2}$.
\subsection{Double binding-energy difference and Belyaev formula \label{sec:DBED}}

\begin{figure}[ht]
\begin{center}
\includegraphics[width=83mm]{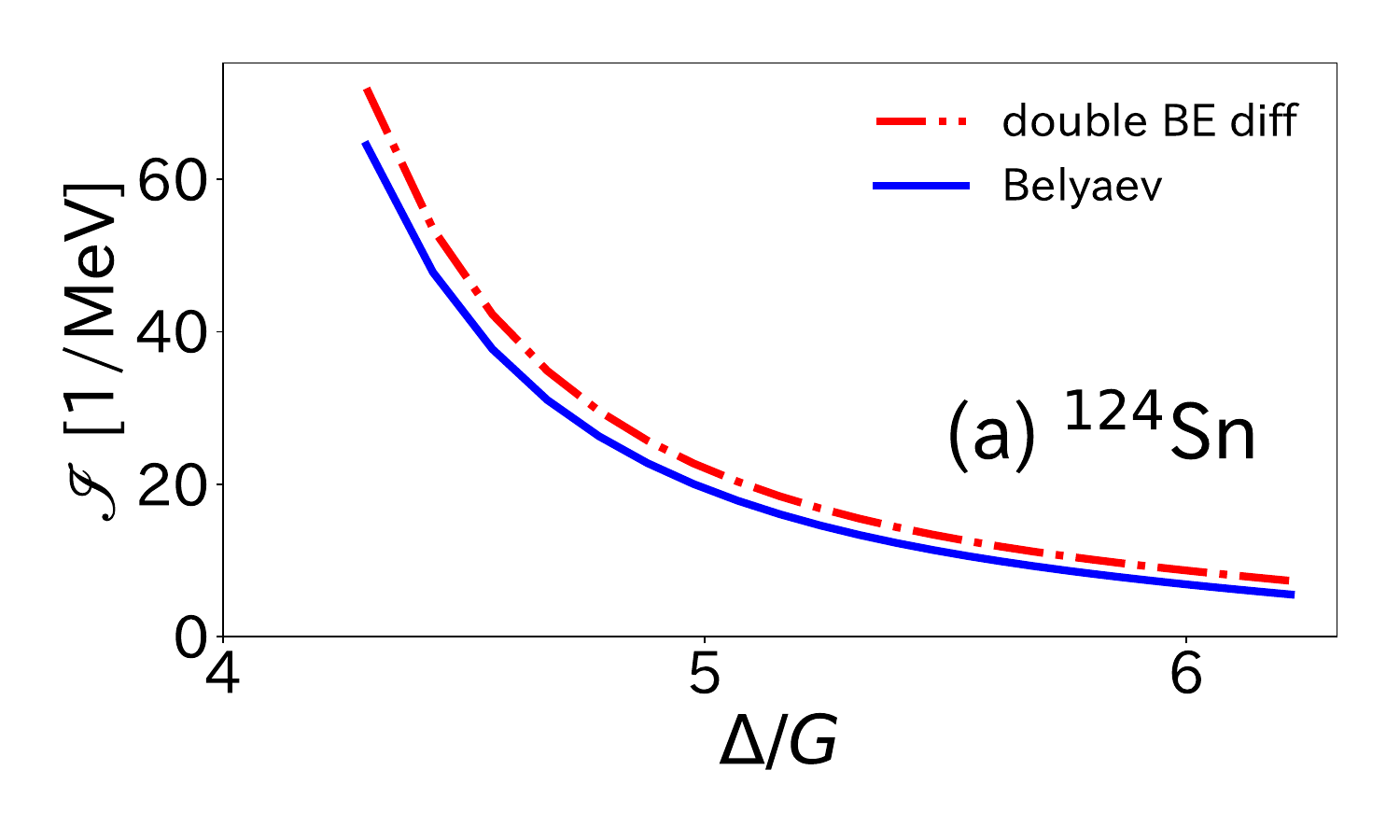}\\
\includegraphics[width=83mm]{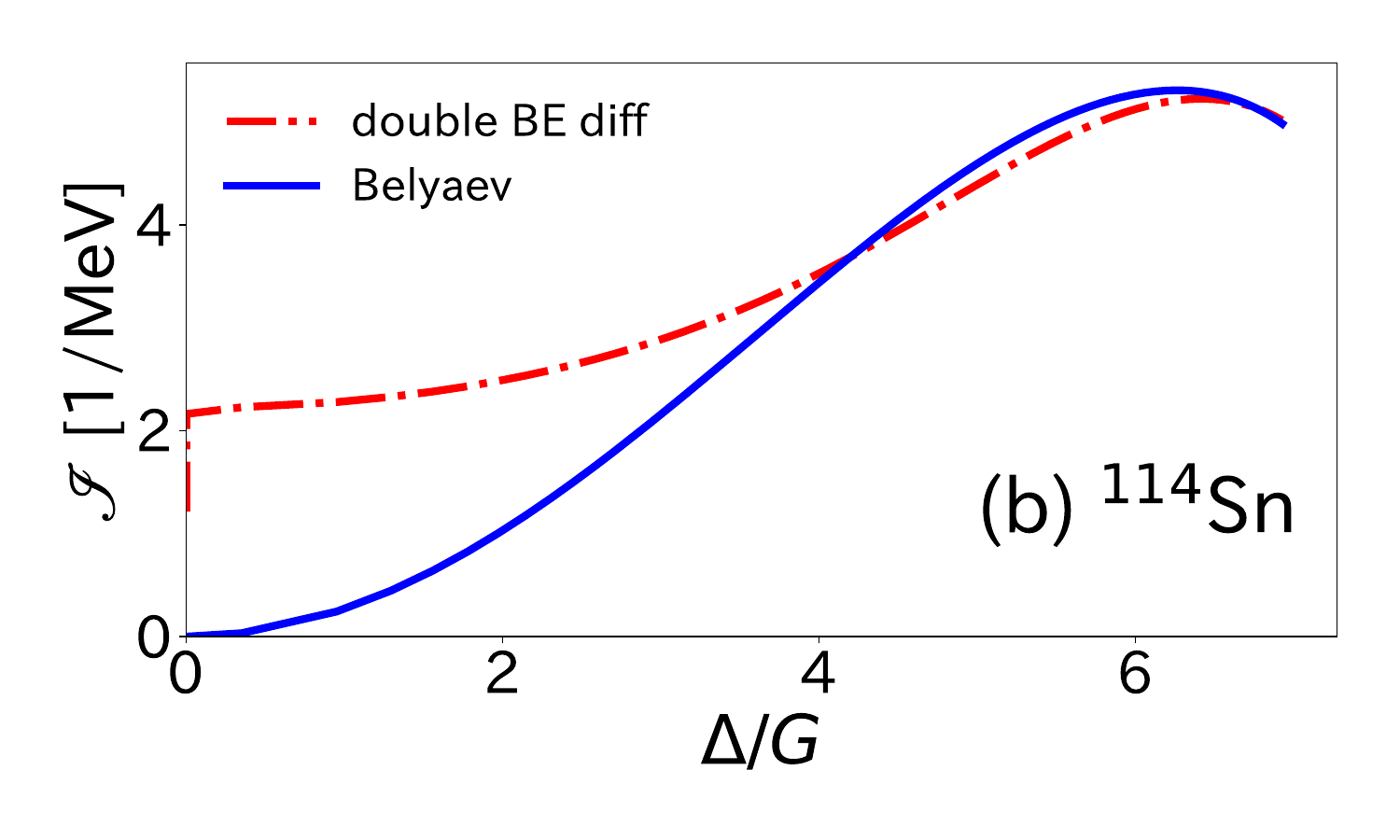}
\end{center}
\caption{Comparison of the P-MoI calculated with the Belyaev inertia (blue curve) and with the double binding-energy difference (red curve) for open-shell $^{124}$Sn (a) and closed-subshell $^{114}$Sn (b) nuclei.
\label{fig:DGMOIcrank}}
\end{figure}

In Fig.~\ref{fig:MOI-N},
the P-MoI calculated with the formula of double binding-energy differences (\ref{eq:pairingMOIapr})
is significantly different from
that of the Belyaev formula (\ref{eq:MOIpair})
near the magic numbers.
This is shown more in detail in Fig.~\ref{fig:DGMOIcrank}.
In an open-shell system such as $^{124}$Sn, the P-MoI's
evaluated with the two formulae
agree in a wide range of the pair amplitude, 
while a large discrepancy is seen in the region of small pair amplitude
in closed-subshell nuclei, such as $^{114}$Sn.
This discrepancy in the closed-subshell configuration
is understood as follows.
The Belyaev inertia increases with $\Delta^2$ as shown
in Eq.~(\ref{eq:Delta0closedshell}).
On the other hand,
if there is a shell gap between the $N$th and $(N+1)$th
single-particle levels,
there is an abrupt change in the structure, particularly
in the pairing property.
The pairing gaps for open-shell nuclei with $N\pm 2$ are much larger
than that in the closed-(sub)shell nucleus with $N$.
This is responsible for the discrepancy.
At large pair amplitude $\Delta/G$,
the two formulae give almost identical values.
For all the Sn isotopes except for the doubly magic nuclei,
the difference is marginal in the realistic region of the pair amplitude $4\lesssim \Delta/G\lesssim 6$.
Thus, the calculated P-MoI's for Sn isotopes except for the doubly magic nuclei represent the property of the pair rotation, rather than
the empirical shell gap.

\section{conclusion \label{sec:conclusion}}
We have studied the fundamental properties of the P-MoI
for proton-magic Ni, Sn, and Pb isotopes,
with the BCS calculation for a simple monopole pairing interaction
and the Skyrme-DFT calculation using different types of the pairing EDF.
The P-MoI depends on the single-particle level density and
the magnitude of the pair amplitude
which is the order parameter for the normal to the superconducting phase transition.
For open-shell nuclei,
there is a negative correlation between the pair amplitude and P-MoI,
which is opposite to our intuitive picture that the MoI in the
spatial rotation is larger for nuclei with larger deformation.
For Ni and Sn isotopes,
the P-MoI in open-shell nuclei generally increases as a function of 
the neutron number between the two neighboring magic numbers.
This is due to the decrease in the pairing force strength and
the increase in the level density near the Fermi energy, as the neutron number increases.
The P-MoI is also sensitive to the position of the high-$j$ intruder orbits
because they are highly degenerate and increase the level density.
The P-MoI in the BCS approximation can be decomposed into contributions from each single-particle orbit.
This orbital decomposition provides useful insights
and confirms the importance of the intruder orbits.

For closed-shell nuclei,
the P-MoI has the property of an inverse of the empirical 
shell gap.
It suddenly drops at the shell gap and is useful to find
a new magic number in the unknown region.
In fact, the neutron empirical shell gap as a function of the proton number was used for confirming doubly magic nature, for example, $^{78}$Ni \cite{Manea2023, Welker2017}.

It was suggested that there may be a second-order quantum phase transition from a deformed shape to a spherical superconducting phase around $N=66$ in Sn isotopes, to explain the behavior of the two-neutron separation energy
and $B(E2;0_1^+\to 2_1^+)$ \cite{Togashi2018}.
The analysis in the present paper based on the pairing Hamiltonian
with pure spherical single-particle levels
can also explain the behavior of the two-neutron separation energy.
Further investigation is desired to clarify the role of
deformation degrees of freedom.

The order-parameter dependence of the P-MoI is also important for the large-amplitude collective dynamics for the pairing degrees of freedom.
A unified description of the pair vibration and the pair rotation
is necessary to investigate the full dynamics of the nuclear pairing.
The pair rotation takes place only in the superconducting phase
with a non-zero order parameter.
However, even if the ground state is in the normal phase
($\Psi=0$),
the pair rotation can play a role during the collective motion moving away from the equilibrium in Fig.~\ref{fig:potential}
where the order parameter is nonzero, $\Psi\neq 0$.
The transitional situations
from the normal to the superconducting phases
\cite{PhysRevLett.96.032501,Clark2008}
requires a treatment of the large-amplitude collective pair dynamics.
A possible approach is the pairing collective Hamiltonian
\cite{Bes1970, Dussel1971, Dussel1972, Bes1973, Perazzo1978, PhysRevC.97.044310,PhysRevC.98.064327}
where the pair-vibrational and pair-rotational inertias as functions
of the order parameters play an essential role in the collective dynamics.
The work in this direction is currently in progress.

\section*{Acknowledgments}
Discussions with  Tomoya Naito, Shinsuke Ota, Kouhei Washiyama, and Kota Yanase are greatly acknowledged.
This work was supported by the JSPS KAKENHI (Grants No. JP19KK0343, No. JP20K03964,
No. JP23K25864,
No. JP25K07312, and No. JP25H00402) and by the JST ERATO (Grant No. JPMJER2304).

\bibliography{pairmoi}

\begin{thebibliography}{48}%
\makeatletter
\providecommand \@ifxundefined [1]{%
 \@ifx{#1\undefined}
}%
\providecommand \@ifnum [1]{%
 \ifnum #1\expandafter \@firstoftwo
 \else \expandafter \@secondoftwo
 \fi
}%
\providecommand \@ifx [1]{%
 \ifx #1\expandafter \@firstoftwo
 \else \expandafter \@secondoftwo
 \fi
}%
\providecommand \natexlab [1]{#1}%
\providecommand \enquote  [1]{``#1''}%
\providecommand \bibnamefont  [1]{#1}%
\providecommand \bibfnamefont [1]{#1}%
\providecommand \citenamefont [1]{#1}%
\providecommand \href@noop [0]{\@secondoftwo}%
\providecommand \href [0]{\begingroup \@sanitize@url \@href}%
\providecommand \@href[1]{\@@startlink{#1}\@@href}%
\providecommand \@@href[1]{\endgroup#1\@@endlink}%
\providecommand \@sanitize@url [0]{\catcode `\\12\catcode `\$12\catcode `\&12\catcode `\#12\catcode `\^12\catcode `\_12\catcode `\%12\relax}%
\providecommand \@@startlink[1]{}%
\providecommand \@@endlink[0]{}%
\providecommand \url  [0]{\begingroup\@sanitize@url \@url }%
\providecommand \@url [1]{\endgroup\@href {#1}{\urlprefix }}%
\providecommand \urlprefix  [0]{URL }%
\providecommand \Eprint [0]{\href }%
\providecommand \doibase [0]{https://doi.org/}%
\providecommand \selectlanguage [0]{\@gobble}%
\providecommand \bibinfo  [0]{\@secondoftwo}%
\providecommand \bibfield  [0]{\@secondoftwo}%
\providecommand \translation [1]{[#1]}%
\providecommand \BibitemOpen [0]{}%
\providecommand \bibitemStop [0]{}%
\providecommand \bibitemNoStop [0]{.\EOS\space}%
\providecommand \EOS [0]{\spacefactor3000\relax}%
\providecommand \BibitemShut  [1]{\csname bibitem#1\endcsname}%
\let\auto@bib@innerbib\@empty
\bibitem [{\citenamefont {Dean}\ and\ \citenamefont {Hjorth-Jensen}(2003)}]{Dean2003}%
  \BibitemOpen
  \bibfield  {author} {\bibinfo {author} {\bibfnamefont {D.~J.}\ \bibnamefont {Dean}}\ and\ \bibinfo {author} {\bibfnamefont {M.}~\bibnamefont {Hjorth-Jensen}},\ }\bibfield  {title} {\bibinfo {title} {Pairing in nuclear systems: from neutron stars to finite nuclei},\ }\href {https://doi.org/10.1103/revmodphys.75.607} {\bibfield  {journal} {\bibinfo  {journal} {Rev. Mod. Phys.}\ }\textbf {\bibinfo {volume} {75}},\ \bibinfo {pages} {607} (\bibinfo {year} {2003})}\BibitemShut {NoStop}%
\bibitem [{\citenamefont {Brink}\ and\ \citenamefont {Broglia}(2005)}]{Brink-Broglia}%
  \BibitemOpen
  \bibfield  {author} {\bibinfo {author} {\bibfnamefont {D.~M.}\ \bibnamefont {Brink}}\ and\ \bibinfo {author} {\bibfnamefont {R.~A.}\ \bibnamefont {Broglia}},\ }\href@noop {} {\emph {\bibinfo {title} {Nuclear Superfluidity, Pairing in Finite Systems}}}\ (\bibinfo  {publisher} {Cambridge University Press, Cambridge, UK},\ \bibinfo {year} {2005})\BibitemShut {NoStop}%
\bibitem [{\citenamefont {Potel}\ \emph {et~al.}(2011)\citenamefont {Potel}, \citenamefont {Barranco}, \citenamefont {Marini}, \citenamefont {Idini}, \citenamefont {Vigezzi},\ and\ \citenamefont {Broglia}}]{Potel2011}%
  \BibitemOpen
  \bibfield  {author} {\bibinfo {author} {\bibfnamefont {G.}~\bibnamefont {Potel}}, \bibinfo {author} {\bibfnamefont {F.}~\bibnamefont {Barranco}}, \bibinfo {author} {\bibfnamefont {F.}~\bibnamefont {Marini}}, \bibinfo {author} {\bibfnamefont {A.}~\bibnamefont {Idini}}, \bibinfo {author} {\bibfnamefont {E.}~\bibnamefont {Vigezzi}},\ and\ \bibinfo {author} {\bibfnamefont {R.~A.}\ \bibnamefont {Broglia}},\ }\bibfield  {title} {\bibinfo {title} {Calculation of the transition from pairing vibrational to pairing rotational regimes between magic nuclei $^{100}${Sn} and $^{132}${Sn} via two-nucleon transfer reactions},\ }\href {https://doi.org/10.1103/physrevlett.107.092501} {\bibfield  {journal} {\bibinfo  {journal} {Phys. Rev. Lett.}\ }\textbf {\bibinfo {volume} {107}},\ \bibinfo {pages} {092501} (\bibinfo {year} {2011})}\BibitemShut {NoStop}%
\bibitem [{\citenamefont {Potel}\ \emph {et~al.}(2013{\natexlab{a}})\citenamefont {Potel}, \citenamefont {Idini}, \citenamefont {Barranco}, \citenamefont {Vigezzi},\ and\ \citenamefont {Broglia}}]{Potel2013}%
  \BibitemOpen
  \bibfield  {author} {\bibinfo {author} {\bibfnamefont {G.}~\bibnamefont {Potel}}, \bibinfo {author} {\bibfnamefont {A.}~\bibnamefont {Idini}}, \bibinfo {author} {\bibfnamefont {F.}~\bibnamefont {Barranco}}, \bibinfo {author} {\bibfnamefont {E.}~\bibnamefont {Vigezzi}},\ and\ \bibinfo {author} {\bibfnamefont {R.~A.}\ \bibnamefont {Broglia}},\ }\bibfield  {title} {\bibinfo {title} {Cooper pair transfer in nuclei},\ }\href {https://doi.org/10.1088/0034-4885/76/10/106301} {\bibfield  {journal} {\bibinfo  {journal} {Rep. Prog. Phys.}\ }\textbf {\bibinfo {volume} {76}},\ \bibinfo {pages} {106301} (\bibinfo {year} {2013}{\natexlab{a}})}\BibitemShut {NoStop}%
\bibitem [{\citenamefont {Potel}\ \emph {et~al.}(2013{\natexlab{b}})\citenamefont {Potel}, \citenamefont {Idini}, \citenamefont {Barranco}, \citenamefont {Vigezzi},\ and\ \citenamefont {Broglia}}]{PhysRevC.87.054321}%
  \BibitemOpen
  \bibfield  {author} {\bibinfo {author} {\bibfnamefont {G.}~\bibnamefont {Potel}}, \bibinfo {author} {\bibfnamefont {A.}~\bibnamefont {Idini}}, \bibinfo {author} {\bibfnamefont {F.}~\bibnamefont {Barranco}}, \bibinfo {author} {\bibfnamefont {E.}~\bibnamefont {Vigezzi}},\ and\ \bibinfo {author} {\bibfnamefont {R.~A.}\ \bibnamefont {Broglia}},\ }\bibfield  {title} {\bibinfo {title} {{Quantitative study of coherent pairing modes with two-neutron transfer: Sn isotopes}},\ }\href {https://doi.org/10.1103/PhysRevC.87.054321} {\bibfield  {journal} {\bibinfo  {journal} {Phys. Rev. C}\ }\textbf {\bibinfo {volume} {87}},\ \bibinfo {pages} {054321} (\bibinfo {year} {2013}{\natexlab{b}})}\BibitemShut {NoStop}%
\bibitem [{\citenamefont {Hagino}\ and\ \citenamefont {Scamps}(2015)}]{PhysRevC.92.064602}%
  \BibitemOpen
  \bibfield  {author} {\bibinfo {author} {\bibfnamefont {K.}~\bibnamefont {Hagino}}\ and\ \bibinfo {author} {\bibfnamefont {G.}~\bibnamefont {Scamps}},\ }\bibfield  {title} {\bibinfo {title} {Enhancement factor for two-neutron transfer reactions with a schematic coupled-channels model},\ }\href {https://doi.org/10.1103/PhysRevC.92.064602} {\bibfield  {journal} {\bibinfo  {journal} {Phys. Rev. C}\ }\textbf {\bibinfo {volume} {92}},\ \bibinfo {pages} {064602} (\bibinfo {year} {2015})}\BibitemShut {NoStop}%
\bibitem [{\citenamefont {Ring}\ and\ \citenamefont {Schuck}(1980)}]{RS80}%
  \BibitemOpen
  \bibfield  {author} {\bibinfo {author} {\bibfnamefont {P.}~\bibnamefont {Ring}}\ and\ \bibinfo {author} {\bibfnamefont {P.}~\bibnamefont {Schuck}},\ }\href@noop {} {\emph {\bibinfo {title} {The nuclear many-body problems}}},\ Texts and monographs in physics\ (\bibinfo  {publisher} {Springer-Verlag},\ \bibinfo {address} {New York},\ \bibinfo {year} {1980})\BibitemShut {NoStop}%
\bibitem [{\citenamefont {Bohr}\ and\ \citenamefont {Mottelson}(1975)}]{BM75}%
  \BibitemOpen
  \bibfield  {author} {\bibinfo {author} {\bibfnamefont {A.}~\bibnamefont {Bohr}}\ and\ \bibinfo {author} {\bibfnamefont {B.~R.}\ \bibnamefont {Mottelson}},\ }\href@noop {} {\emph {\bibinfo {title} {Nuclear Structure, Vol. II}}}\ (\bibinfo  {publisher} {W. A. Benjamin},\ \bibinfo {address} {New York},\ \bibinfo {year} {1975})\BibitemShut {NoStop}%
\bibitem [{\citenamefont {Bohr}\ \emph {et~al.}(1958)\citenamefont {Bohr}, \citenamefont {Mottelson},\ and\ \citenamefont {Pines}}]{BMP58}%
  \BibitemOpen
  \bibfield  {author} {\bibinfo {author} {\bibfnamefont {A.}~\bibnamefont {Bohr}}, \bibinfo {author} {\bibfnamefont {B.~R.}\ \bibnamefont {Mottelson}},\ and\ \bibinfo {author} {\bibfnamefont {D.}~\bibnamefont {Pines}},\ }\bibfield  {title} {\bibinfo {title} {Possible analogy between the excitation spectra of nuclei and those of the superconducting metallic state},\ }\href {https://doi.org/10.1103/PhysRev.110.936} {\bibfield  {journal} {\bibinfo  {journal} {Phys. Rev.}\ }\textbf {\bibinfo {volume} {110}},\ \bibinfo {pages} {936} (\bibinfo {year} {1958})}\BibitemShut {NoStop}%
\bibitem [{\citenamefont {Broglia}\ \emph {et~al.}(2000)\citenamefont {Broglia}, \citenamefont {Terasaki},\ and\ \citenamefont {Giovanardi}}]{Broglia2000}%
  \BibitemOpen
  \bibfield  {author} {\bibinfo {author} {\bibfnamefont {R.}~\bibnamefont {Broglia}}, \bibinfo {author} {\bibfnamefont {J.}~\bibnamefont {Terasaki}},\ and\ \bibinfo {author} {\bibfnamefont {N.}~\bibnamefont {Giovanardi}},\ }\bibfield  {title} {\bibinfo {title} {The {Anderson–Goldstone–Nambu} mode in finite and in infinite systems},\ }\href {https://doi.org/10.1016/s0370-1573(00)00046-6} {\bibfield  {journal} {\bibinfo  {journal} {Phys. Rep.}\ }\textbf {\bibinfo {volume} {335}},\ \bibinfo {pages} {1} (\bibinfo {year} {2000})}\BibitemShut {NoStop}%
\bibitem [{\citenamefont {Anderson}(1958)}]{PhysRev.112.1900}%
  \BibitemOpen
  \bibfield  {author} {\bibinfo {author} {\bibfnamefont {P.~W.}\ \bibnamefont {Anderson}},\ }\bibfield  {title} {\bibinfo {title} {Random-phase approximation in the theory of superconductivity},\ }\href {https://doi.org/10.1103/PhysRev.112.1900} {\bibfield  {journal} {\bibinfo  {journal} {Phys. Rev.}\ }\textbf {\bibinfo {volume} {112}},\ \bibinfo {pages} {1900} (\bibinfo {year} {1958})}\BibitemShut {NoStop}%
\bibitem [{\citenamefont {Nambu}(1960)}]{PhysRev.117.648}%
  \BibitemOpen
  \bibfield  {author} {\bibinfo {author} {\bibfnamefont {Y.}~\bibnamefont {Nambu}},\ }\bibfield  {title} {\bibinfo {title} {Quasi-particles and gauge invariance in the theory of superconductivity},\ }\href {https://doi.org/10.1103/PhysRev.117.648} {\bibfield  {journal} {\bibinfo  {journal} {Phys. Rev.}\ }\textbf {\bibinfo {volume} {117}},\ \bibinfo {pages} {648} (\bibinfo {year} {1960})}\BibitemShut {NoStop}%
\bibitem [{\citenamefont {Goldstone}(1961)}]{INC_19_154}%
  \BibitemOpen
  \bibfield  {author} {\bibinfo {author} {\bibfnamefont {J.}~\bibnamefont {Goldstone}},\ }\bibfield  {title} {\bibinfo {title} {Field theories with superconductor solutions},\ }\href {https://doi.org/10.1007/BF02812722} {\bibfield  {journal} {\bibinfo  {journal} {Il Nuovo Cimento}\ }\textbf {\bibinfo {volume} {19}},\ \bibinfo {pages} {154} (\bibinfo {year} {1961})}\BibitemShut {NoStop}%
\bibitem [{\citenamefont {Papenbrock}(2022)}]{Papenbrock2022}%
  \BibitemOpen
  \bibfield  {author} {\bibinfo {author} {\bibfnamefont {T.}~\bibnamefont {Papenbrock}},\ }\bibfield  {title} {\bibinfo {title} {Effective field theory of pairing rotations},\ }\href {https://doi.org/10.1103/physrevc.105.044322} {\bibfield  {journal} {\bibinfo  {journal} {Phys. Rev. C}\ }\textbf {\bibinfo {volume} {105}},\ \bibinfo {pages} {044322} (\bibinfo {year} {2022})}\BibitemShut {NoStop}%
\bibitem [{\citenamefont {Bender}\ \emph {et~al.}(2000)\citenamefont {Bender}, \citenamefont {Rutz}, \citenamefont {Reinhard},\ and\ \citenamefont {Maruhn}}]{Bender2000}%
  \BibitemOpen
  \bibfield  {author} {\bibinfo {author} {\bibfnamefont {M.}~\bibnamefont {Bender}}, \bibinfo {author} {\bibfnamefont {K.}~\bibnamefont {Rutz}}, \bibinfo {author} {\bibfnamefont {P.~G.}\ \bibnamefont {Reinhard}},\ and\ \bibinfo {author} {\bibfnamefont {J.~A.}\ \bibnamefont {Maruhn}},\ }\bibfield  {title} {\bibinfo {title} {Pairing gaps from nuclear mean-field models},\ }\href {https://doi.org/10.1007/s10050-000-4504-z} {\bibfield  {journal} {\bibinfo  {journal} {Eur. Phys. J. A}\ }\textbf {\bibinfo {volume} {8}},\ \bibinfo {pages} {59} (\bibinfo {year} {2000})}\BibitemShut {NoStop}%
\bibitem [{\citenamefont {Bertsch}\ \emph {et~al.}(2009)\citenamefont {Bertsch}, \citenamefont {Bertulani}, \citenamefont {Nazarewicz}, \citenamefont {Schunck},\ and\ \citenamefont {Stoitsov}}]{Bertsch2009}%
  \BibitemOpen
  \bibfield  {author} {\bibinfo {author} {\bibfnamefont {G.~F.}\ \bibnamefont {Bertsch}}, \bibinfo {author} {\bibfnamefont {C.~A.}\ \bibnamefont {Bertulani}}, \bibinfo {author} {\bibfnamefont {W.}~\bibnamefont {Nazarewicz}}, \bibinfo {author} {\bibfnamefont {N.}~\bibnamefont {Schunck}},\ and\ \bibinfo {author} {\bibfnamefont {M.~V.}\ \bibnamefont {Stoitsov}},\ }\bibfield  {title} {\bibinfo {title} {Odd-even mass differences from self-consistent mean field theory},\ }\href {https://doi.org/10.1103/physrevc.79.034306} {\bibfield  {journal} {\bibinfo  {journal} {Phys. Rev. C}\ }\textbf {\bibinfo {volume} {79}},\ \bibinfo {pages} {034306} (\bibinfo {year} {2009})}\BibitemShut {NoStop}%
\bibitem [{\citenamefont {Schunck}\ \emph {et~al.}(2010)\citenamefont {Schunck}, \citenamefont {Dobaczewski}, \citenamefont {McDonnell}, \citenamefont {Moré}, \citenamefont {Nazarewicz}, \citenamefont {Sarich},\ and\ \citenamefont {Stoitsov}}]{Schunck2010}%
  \BibitemOpen
  \bibfield  {author} {\bibinfo {author} {\bibfnamefont {N.}~\bibnamefont {Schunck}}, \bibinfo {author} {\bibfnamefont {J.}~\bibnamefont {Dobaczewski}}, \bibinfo {author} {\bibfnamefont {J.}~\bibnamefont {McDonnell}}, \bibinfo {author} {\bibfnamefont {J.}~\bibnamefont {Moré}}, \bibinfo {author} {\bibfnamefont {W.}~\bibnamefont {Nazarewicz}}, \bibinfo {author} {\bibfnamefont {J.}~\bibnamefont {Sarich}},\ and\ \bibinfo {author} {\bibfnamefont {M.~V.}\ \bibnamefont {Stoitsov}},\ }\bibfield  {title} {\bibinfo {title} {One-quasiparticle states in the nuclear energy density functional theory},\ }\href {https://doi.org/10.1103/physrevc.81.024316} {\bibfield  {journal} {\bibinfo  {journal} {Phys. Rev. C}\ }\textbf {\bibinfo {volume} {81}},\ \bibinfo {pages} {024316} (\bibinfo {year} {2010})}\BibitemShut {NoStop}%
\bibitem [{\citenamefont {Beck}\ \emph {et~al.}(1972)\citenamefont {Beck}, \citenamefont {Kleber},\ and\ \citenamefont {Schmidt}}]{ZPA250_155}%
  \BibitemOpen
  \bibfield  {author} {\bibinfo {author} {\bibfnamefont {R.}~\bibnamefont {Beck}}, \bibinfo {author} {\bibfnamefont {M.}~\bibnamefont {Kleber}},\ and\ \bibinfo {author} {\bibfnamefont {H.}~\bibnamefont {Schmidt}},\ }\bibfield  {title} {\bibinfo {title} {Pairing rotations and separation energies},\ }\href {https://doi.org/10.1007/BF01386946} {\bibfield  {journal} {\bibinfo  {journal} {Z. Phys.}\ }\textbf {\bibinfo {volume} {250}},\ \bibinfo {pages} {155 } (\bibinfo {year} {1972})}\BibitemShut {NoStop}%
\bibitem [{\citenamefont {Krappe}(1975)}]{Krappe1975}%
  \BibitemOpen
  \bibfield  {author} {\bibinfo {author} {\bibfnamefont {H.~J.}\ \bibnamefont {Krappe}},\ }\bibfield  {title} {\bibinfo {title} {On the use of a variable moment of pairing},\ }\href {https://doi.org/10.1007/bf01409299} {\bibfield  {journal} {\bibinfo  {journal} {Z. Phys. A}\ }\textbf {\bibinfo {volume} {275}},\ \bibinfo {pages} {297} (\bibinfo {year} {1975})}\BibitemShut {NoStop}%
\bibitem [{\citenamefont {Hinohara}\ and\ \citenamefont {Nazarewicz}(2016)}]{Hinohara2016}%
  \BibitemOpen
  \bibfield  {author} {\bibinfo {author} {\bibfnamefont {N.}~\bibnamefont {Hinohara}}\ and\ \bibinfo {author} {\bibfnamefont {W.}~\bibnamefont {Nazarewicz}},\ }\bibfield  {title} {\bibinfo {title} {Pairing {Nambu-Goldstone} modes within nuclear density functional theory},\ }\href {https://doi.org/10.1103/physrevlett.116.152502} {\bibfield  {journal} {\bibinfo  {journal} {Phys. Rev. Lett.}\ }\textbf {\bibinfo {volume} {116}},\ \bibinfo {pages} {152502} (\bibinfo {year} {2016})}\BibitemShut {NoStop}%
\bibitem [{\citenamefont {Hinohara}(2018)}]{Hinohara2018}%
  \BibitemOpen
  \bibfield  {author} {\bibinfo {author} {\bibfnamefont {N.}~\bibnamefont {Hinohara}},\ }\bibfield  {title} {\bibinfo {title} {Extending pairing energy density functional using pairing rotational moments of inertia},\ }\href {https://doi.org/10.1088/1361-6471/aa9f8b} {\bibfield  {journal} {\bibinfo  {journal} {J. Phys. G}\ }\textbf {\bibinfo {volume} {45}},\ \bibinfo {pages} {024004} (\bibinfo {year} {2018})}\BibitemShut {NoStop}%
\bibitem [{\citenamefont {Reinhard}\ and\ \citenamefont {Nazarewicz}(2017)}]{Reinhard2017}%
  \BibitemOpen
  \bibfield  {author} {\bibinfo {author} {\bibfnamefont {P.-G.}\ \bibnamefont {Reinhard}}\ and\ \bibinfo {author} {\bibfnamefont {W.}~\bibnamefont {Nazarewicz}},\ }\bibfield  {title} {\bibinfo {title} {Toward a global description of nuclear charge radii: Exploring the {Fayans} energy density functional},\ }\href {https://doi.org/10.1103/physrevc.95.064328} {\bibfield  {journal} {\bibinfo  {journal} {Phys. Rev. C}\ }\textbf {\bibinfo {volume} {95}},\ \bibinfo {pages} {064328} (\bibinfo {year} {2017})}\BibitemShut {NoStop}%
\bibitem [{\citenamefont {Reinhard}\ \emph {et~al.}(2024)\citenamefont {Reinhard}, \citenamefont {O’Neal}, \citenamefont {Wild},\ and\ \citenamefont {Nazarewicz}}]{Reinhard_2024}%
  \BibitemOpen
  \bibfield  {author} {\bibinfo {author} {\bibfnamefont {P.-G.}\ \bibnamefont {Reinhard}}, \bibinfo {author} {\bibfnamefont {J.}~\bibnamefont {O’Neal}}, \bibinfo {author} {\bibfnamefont {S.~M.}\ \bibnamefont {Wild}},\ and\ \bibinfo {author} {\bibfnamefont {W.}~\bibnamefont {Nazarewicz}},\ }\bibfield  {title} {\bibinfo {title} {Extended fayans energy density functional: optimization and analysis},\ }\href {https://doi.org/10.1088/1361-6471/ad633a} {\bibfield  {journal} {\bibinfo  {journal} {J. Phys. G: Nucl. Part. Phys.}\ }\textbf {\bibinfo {volume} {51}},\ \bibinfo {pages} {105101} (\bibinfo {year} {2024})}\BibitemShut {NoStop}%
\bibitem [{\citenamefont {Kouno}\ \emph {et~al.}(2021)\citenamefont {Kouno}, \citenamefont {Ishizuka}, \citenamefont {Inakura},\ and\ \citenamefont {Chiba}}]{Kouno2021}%
  \BibitemOpen
  \bibfield  {author} {\bibinfo {author} {\bibfnamefont {T.}~\bibnamefont {Kouno}}, \bibinfo {author} {\bibfnamefont {C.}~\bibnamefont {Ishizuka}}, \bibinfo {author} {\bibfnamefont {T.}~\bibnamefont {Inakura}},\ and\ \bibinfo {author} {\bibfnamefont {S.}~\bibnamefont {Chiba}},\ }\bibfield  {title} {\bibinfo {title} {Pairing strength in the relativistic mean-field theory determined from the fission barrier heights of actinide nuclei and verified by pairing rotation and binding energies},\ }\href {https://doi.org/10.1093/ptep/ptab167} {\bibfield  {journal} {\bibinfo  {journal} {Prog. Theor. Exp. Phys.}\ }\textbf {\bibinfo {volume} {2022}},\ \bibinfo {pages} {023D02} (\bibinfo {year} {2021})}\BibitemShut {NoStop}%
\bibitem [{\citenamefont {Wen}\ and\ \citenamefont {Nakatsukasa}(2022)}]{WN22}%
  \BibitemOpen
  \bibfield  {author} {\bibinfo {author} {\bibfnamefont {K.}~\bibnamefont {Wen}}\ and\ \bibinfo {author} {\bibfnamefont {T.}~\bibnamefont {Nakatsukasa}},\ }\bibfield  {title} {\bibinfo {title} {Microscopic collective inertial masses for nuclear reaction in the presence of nucleonic effective mass},\ }\href {https://doi.org/10.1103/PhysRevC.105.034603} {\bibfield  {journal} {\bibinfo  {journal} {Phys. Rev. C}\ }\textbf {\bibinfo {volume} {105}},\ \bibinfo {pages} {034603} (\bibinfo {year} {2022})}\BibitemShut {NoStop}%
\bibitem [{\citenamefont {Lunney}\ \emph {et~al.}(2003)\citenamefont {Lunney}, \citenamefont {Pearson},\ and\ \citenamefont {Thibault}}]{Lunney2003}%
  \BibitemOpen
  \bibfield  {author} {\bibinfo {author} {\bibfnamefont {D.}~\bibnamefont {Lunney}}, \bibinfo {author} {\bibfnamefont {J.~M.}\ \bibnamefont {Pearson}},\ and\ \bibinfo {author} {\bibfnamefont {C.}~\bibnamefont {Thibault}},\ }\bibfield  {title} {\bibinfo {title} {Recent trends in the determination of nuclear masses},\ }\href {https://doi.org/10.1103/revmodphys.75.1021} {\bibfield  {journal} {\bibinfo  {journal} {Rev. Mod. Phys.}\ }\textbf {\bibinfo {volume} {75}},\ \bibinfo {pages} {1021} (\bibinfo {year} {2003})}\BibitemShut {NoStop}%
\bibitem [{\citenamefont {Stoitsov}\ \emph {et~al.}(2005)\citenamefont {Stoitsov}, \citenamefont {Dobaczewski}, \citenamefont {Nazarewicz},\ and\ \citenamefont {Ring}}]{Stoitsov2005}%
  \BibitemOpen
  \bibfield  {author} {\bibinfo {author} {\bibfnamefont {M.}~\bibnamefont {Stoitsov}}, \bibinfo {author} {\bibfnamefont {J.}~\bibnamefont {Dobaczewski}}, \bibinfo {author} {\bibfnamefont {W.}~\bibnamefont {Nazarewicz}},\ and\ \bibinfo {author} {\bibfnamefont {P.}~\bibnamefont {Ring}},\ }\bibfield  {title} {\bibinfo {title} {Axially deformed solution of the {Skyrme–Hartree–Fock–Bogolyubov} equations using the transformed harmonic oscillator basis. {The} program {HFBTHO} (v1.66p)},\ }\href {https://doi.org/10.1016/j.cpc.2005.01.001} {\bibfield  {journal} {\bibinfo  {journal} {Comp. Phys. Commun.}\ }\textbf {\bibinfo {volume} {167}},\ \bibinfo {pages} {43} (\bibinfo {year} {2005})}\BibitemShut {NoStop}%
\bibitem [{\citenamefont {Vautherin}\ and\ \citenamefont {Brink}(1972)}]{Vautherin1972}%
  \BibitemOpen
  \bibfield  {author} {\bibinfo {author} {\bibfnamefont {D.}~\bibnamefont {Vautherin}}\ and\ \bibinfo {author} {\bibfnamefont {D.~M.}\ \bibnamefont {Brink}},\ }\bibfield  {title} {\bibinfo {title} {{Hartree-Fock} calculations with {Skyrme’s} interaction. {I.} {Spherical} nuclei},\ }\href {https://doi.org/10.1103/physrevc.5.626} {\bibfield  {journal} {\bibinfo  {journal} {Phys. Rev. C}\ }\textbf {\bibinfo {volume} {5}},\ \bibinfo {pages} {626} (\bibinfo {year} {1972})}\BibitemShut {NoStop}%
\bibitem [{\citenamefont {Bender}\ \emph {et~al.}(2003)\citenamefont {Bender}, \citenamefont {Heenen},\ and\ \citenamefont {Reinhard}}]{Bender2003}%
  \BibitemOpen
  \bibfield  {author} {\bibinfo {author} {\bibfnamefont {M.}~\bibnamefont {Bender}}, \bibinfo {author} {\bibfnamefont {P.-H.}\ \bibnamefont {Heenen}},\ and\ \bibinfo {author} {\bibfnamefont {P.-G.}\ \bibnamefont {Reinhard}},\ }\bibfield  {title} {\bibinfo {title} {Self-consistent mean-field models for nuclear structure},\ }\href {https://doi.org/10.1103/revmodphys.75.121} {\bibfield  {journal} {\bibinfo  {journal} {Rev. Mod. Phys.}\ }\textbf {\bibinfo {volume} {75}},\ \bibinfo {pages} {121} (\bibinfo {year} {2003})}\BibitemShut {NoStop}%
\bibitem [{\citenamefont {Nilsson}\ and\ \citenamefont {Ragnarsson}(2005)}]{Nilsson}%
  \BibitemOpen
  \bibfield  {author} {\bibinfo {author} {\bibfnamefont {S.~G.}\ \bibnamefont {Nilsson}}\ and\ \bibinfo {author} {\bibfnamefont {I.}~\bibnamefont {Ragnarsson}},\ }\href@noop {} {\emph {\bibinfo {title} {Shapes and Shells in Nuclear Structure}}}\ (\bibinfo  {publisher} {Cambridge University Press},\ \bibinfo {year} {2005})\BibitemShut {NoStop}%
\bibitem [{\citenamefont {Chabanat}\ \emph {et~al.}(1998)\citenamefont {Chabanat}, \citenamefont {Bonche}, \citenamefont {Haensel}, \citenamefont {Meyer},\ and\ \citenamefont {Schaeffer}}]{Chabanat1998}%
  \BibitemOpen
  \bibfield  {author} {\bibinfo {author} {\bibfnamefont {E.}~\bibnamefont {Chabanat}}, \bibinfo {author} {\bibfnamefont {P.}~\bibnamefont {Bonche}}, \bibinfo {author} {\bibfnamefont {P.}~\bibnamefont {Haensel}}, \bibinfo {author} {\bibfnamefont {J.}~\bibnamefont {Meyer}},\ and\ \bibinfo {author} {\bibfnamefont {R.}~\bibnamefont {Schaeffer}},\ }\bibfield  {title} {\bibinfo {title} {A {Skyrme} parametrization from subnuclear to neutron star densities {Part II}. {Nuclei} far from stabilities},\ }\href {https://doi.org/10.1016/s0375-9474(98)00180-8} {\bibfield  {journal} {\bibinfo  {journal} {Nucl. Phys. A}\ }\textbf {\bibinfo {volume} {635}},\ \bibinfo {pages} {231} (\bibinfo {year} {1998})}\BibitemShut {NoStop}%
\bibitem [{\citenamefont {Stoitsov}\ \emph {et~al.}(2013)\citenamefont {Stoitsov}, \citenamefont {Schunck}, \citenamefont {Kortelainen}, \citenamefont {Michel}, \citenamefont {Nam}, \citenamefont {Olsen}, \citenamefont {Sarich},\ and\ \citenamefont {Wild}}]{Stoitsov2013}%
  \BibitemOpen
  \bibfield  {author} {\bibinfo {author} {\bibfnamefont {M.}~\bibnamefont {Stoitsov}}, \bibinfo {author} {\bibfnamefont {N.}~\bibnamefont {Schunck}}, \bibinfo {author} {\bibfnamefont {M.}~\bibnamefont {Kortelainen}}, \bibinfo {author} {\bibfnamefont {N.}~\bibnamefont {Michel}}, \bibinfo {author} {\bibfnamefont {H.}~\bibnamefont {Nam}}, \bibinfo {author} {\bibfnamefont {E.}~\bibnamefont {Olsen}}, \bibinfo {author} {\bibfnamefont {J.}~\bibnamefont {Sarich}},\ and\ \bibinfo {author} {\bibfnamefont {S.}~\bibnamefont {Wild}},\ }\bibfield  {title} {\bibinfo {title} {Axially deformed solution of the {Skyrme-Hartree–Fock–Bogoliubov} equations using the transformed harmonic oscillator basis ({II}) {\sc hfbtho} v2.00d: A new version of the program},\ }\href {https://doi.org/10.1016/j.cpc.2013.01.013} {\bibfield  {journal} {\bibinfo  {journal} {Comp. Phys. Commun.}\ }\textbf {\bibinfo {volume} {184}},\ \bibinfo {pages} {1592} (\bibinfo {year} {2013})}\BibitemShut {NoStop}%
\bibitem [{\citenamefont {Wang}\ \emph {et~al.}(2021)\citenamefont {Wang}, \citenamefont {Huang}, \citenamefont {Kondev}, \citenamefont {Audi},\ and\ \citenamefont {Naimi}}]{Wang2021}%
  \BibitemOpen
  \bibfield  {author} {\bibinfo {author} {\bibfnamefont {M.}~\bibnamefont {Wang}}, \bibinfo {author} {\bibfnamefont {W.}~\bibnamefont {Huang}}, \bibinfo {author} {\bibfnamefont {F.}~\bibnamefont {Kondev}}, \bibinfo {author} {\bibfnamefont {G.}~\bibnamefont {Audi}},\ and\ \bibinfo {author} {\bibfnamefont {S.}~\bibnamefont {Naimi}},\ }\bibfield  {title} {\bibinfo {title} {The {AME} 2020 atomic mass evaluation ({II}). {Tables}, graphs and references},\ }\href {https://doi.org/10.1088/1674-1137/abddaf} {\bibfield  {journal} {\bibinfo  {journal} {Chin. Phys. C}\ }\textbf {\bibinfo {volume} {45}},\ \bibinfo {pages} {030003} (\bibinfo {year} {2021})}\BibitemShut {NoStop}%
\bibitem [{\citenamefont {Manea}\ \emph {et~al.}(2023)\citenamefont {Manea}, \citenamefont {Mougeot},\ and\ \citenamefont {Lunney}}]{Manea2023}%
  \BibitemOpen
  \bibfield  {author} {\bibinfo {author} {\bibfnamefont {V.}~\bibnamefont {Manea}}, \bibinfo {author} {\bibfnamefont {M.}~\bibnamefont {Mougeot}},\ and\ \bibinfo {author} {\bibfnamefont {D.}~\bibnamefont {Lunney}},\ }\bibfield  {title} {\bibinfo {title} {The empirical shell gap revisited in light of recent high precision mass spectrometry data},\ }\href {https://doi.org/10.1140/epja/s10050-023-00929-5} {\bibfield  {journal} {\bibinfo  {journal} {Eur. Phys. J. A}\ }\textbf {\bibinfo {volume} {59}},\ \bibinfo {pages} {22} (\bibinfo {year} {2023})}\BibitemShut {NoStop}%
\bibitem [{\citenamefont {Welker}\ \emph {et~al.}(2017)\citenamefont {Welker}, \citenamefont {Althubiti}, \citenamefont {Atanasov}, \citenamefont {Blaum}, \citenamefont {Cocolios}, \citenamefont {Herfurth}, \citenamefont {Kreim}, \citenamefont {Lunney}, \citenamefont {Manea}, \citenamefont {Mougeot}, \citenamefont {Neidherr}, \citenamefont {Nowacki}, \citenamefont {Poves}, \citenamefont {Rosenbusch}, \citenamefont {Schweikhard}, \citenamefont {Wienholtz}, \citenamefont {Wolf},\ and\ \citenamefont {Zuber}}]{Welker2017}%
  \BibitemOpen
  \bibfield  {author} {\bibinfo {author} {\bibfnamefont {A.}~\bibnamefont {Welker}}, \bibinfo {author} {\bibfnamefont {N.~A.~S.}\ \bibnamefont {Althubiti}}, \bibinfo {author} {\bibfnamefont {D.}~\bibnamefont {Atanasov}}, \bibinfo {author} {\bibfnamefont {K.}~\bibnamefont {Blaum}}, \bibinfo {author} {\bibfnamefont {T.~E.}\ \bibnamefont {Cocolios}}, \bibinfo {author} {\bibfnamefont {F.}~\bibnamefont {Herfurth}}, \bibinfo {author} {\bibfnamefont {S.}~\bibnamefont {Kreim}}, \bibinfo {author} {\bibfnamefont {D.}~\bibnamefont {Lunney}}, \bibinfo {author} {\bibfnamefont {V.}~\bibnamefont {Manea}}, \bibinfo {author} {\bibfnamefont {M.}~\bibnamefont {Mougeot}}, \bibinfo {author} {\bibfnamefont {D.}~\bibnamefont {Neidherr}}, \bibinfo {author} {\bibfnamefont {F.}~\bibnamefont {Nowacki}}, \bibinfo {author} {\bibfnamefont {A.}~\bibnamefont {Poves}}, \bibinfo {author} {\bibfnamefont {M.}~\bibnamefont {Rosenbusch}}, \bibinfo {author} {\bibfnamefont {L.}~\bibnamefont {Schweikhard}}, \bibinfo {author} {\bibfnamefont
  {F.}~\bibnamefont {Wienholtz}}, \bibinfo {author} {\bibfnamefont {R.~N.}\ \bibnamefont {Wolf}},\ and\ \bibinfo {author} {\bibfnamefont {K.}~\bibnamefont {Zuber}},\ }\bibfield  {title} {\bibinfo {title} {Binding energy of $^{79}${Cu} : Probing the structure of the doubly magic $^{78}${Ni} from only one proton away},\ }\href {https://doi.org/10.1103/physrevlett.119.192502} {\bibfield  {journal} {\bibinfo  {journal} {Phys. Rev. Lett.}\ }\textbf {\bibinfo {volume} {119}},\ \bibinfo {pages} {192502} (\bibinfo {year} {2017})}\BibitemShut {NoStop}%
\bibitem [{\citenamefont {Ruike}\ \emph {et~al.}(2024)\citenamefont {Ruike}, \citenamefont {Wen}, \citenamefont {Hinohara},\ and\ \citenamefont {Nakatsukasa}}]{EPJWebConf.306.01006}%
  \BibitemOpen
  \bibfield  {author} {\bibinfo {author} {\bibfnamefont {C.}~\bibnamefont {Ruike}}, \bibinfo {author} {\bibfnamefont {K.}~\bibnamefont {Wen}}, \bibinfo {author} {\bibfnamefont {N.}~\bibnamefont {Hinohara}},\ and\ \bibinfo {author} {\bibfnamefont {T.}~\bibnamefont {Nakatsukasa}},\ }\bibfield  {title} {\bibinfo {title} {Collective-subspace requantization for sub-barrier fusion reactions: Inertial functions for collective motions},\ }\href {https://doi.org/10.1051/epjconf/202430601006} {\bibfield  {journal} {\bibinfo  {journal} {EPJ Web Conf.}\ }\textbf {\bibinfo {volume} {306}},\ \bibinfo {pages} {01006} (\bibinfo {year} {2024})}\BibitemShut {NoStop}%
\bibitem [{\citenamefont {Inglis}(1956)}]{Inglis1956}%
  \BibitemOpen
  \bibfield  {author} {\bibinfo {author} {\bibfnamefont {D.~R.}\ \bibnamefont {Inglis}},\ }\bibfield  {title} {\bibinfo {title} {Nuclear moments of inertia due to nucleon motion in a rotating well},\ }\href {https://doi.org/10.1103/physrev.103.1786} {\bibfield  {journal} {\bibinfo  {journal} {Phys. Rev.}\ }\textbf {\bibinfo {volume} {103}},\ \bibinfo {pages} {1786} (\bibinfo {year} {1956})}\BibitemShut {NoStop}%
\bibitem [{\citenamefont {Belyaev}(1965)}]{Belyaev1965}%
  \BibitemOpen
  \bibfield  {author} {\bibinfo {author} {\bibfnamefont {S.~T.}\ \bibnamefont {Belyaev}},\ }\bibfield  {title} {\bibinfo {title} {Time-dependent self-consistent field and collective nuclear {Hamiltonian}},\ }\href {https://doi.org/10.1016/0029-5582(65)90840-0} {\bibfield  {journal} {\bibinfo  {journal} {Nucl. Phys.}\ }\textbf {\bibinfo {volume} {64}},\ \bibinfo {pages} {17} (\bibinfo {year} {1965})}\BibitemShut {NoStop}%
\bibitem [{\citenamefont {Togashi}\ \emph {et~al.}(2018)\citenamefont {Togashi}, \citenamefont {Tsunoda}, \citenamefont {Otsuka}, \citenamefont {Shimizu},\ and\ \citenamefont {Honma}}]{Togashi2018}%
  \BibitemOpen
  \bibfield  {author} {\bibinfo {author} {\bibfnamefont {T.}~\bibnamefont {Togashi}}, \bibinfo {author} {\bibfnamefont {Y.}~\bibnamefont {Tsunoda}}, \bibinfo {author} {\bibfnamefont {T.}~\bibnamefont {Otsuka}}, \bibinfo {author} {\bibfnamefont {N.}~\bibnamefont {Shimizu}},\ and\ \bibinfo {author} {\bibfnamefont {M.}~\bibnamefont {Honma}},\ }\bibfield  {title} {\bibinfo {title} {Novel shape evolution in {Sn} isotopes from magic numbers 50 to 82},\ }\href {https://doi.org/10.1103/physrevlett.121.062501} {\bibfield  {journal} {\bibinfo  {journal} {Phys. Rev. Lett.}\ }\textbf {\bibinfo {volume} {121}},\ \bibinfo {pages} {062501} (\bibinfo {year} {2018})}\BibitemShut {NoStop}%
\bibitem [{\citenamefont {Clark}\ \emph {et~al.}(2006)\citenamefont {Clark}, \citenamefont {Macchiavelli}, \citenamefont {Fortunato},\ and\ \citenamefont {Kr\"ucken}}]{PhysRevLett.96.032501}%
  \BibitemOpen
  \bibfield  {author} {\bibinfo {author} {\bibfnamefont {R.~M.}\ \bibnamefont {Clark}}, \bibinfo {author} {\bibfnamefont {A.~O.}\ \bibnamefont {Macchiavelli}}, \bibinfo {author} {\bibfnamefont {L.}~\bibnamefont {Fortunato}},\ and\ \bibinfo {author} {\bibfnamefont {R.}~\bibnamefont {Kr\"ucken}},\ }\bibfield  {title} {\bibinfo {title} {Critical-point description of the transition from vibrational to rotational regimes in the pairing phase},\ }\href {https://doi.org/10.1103/PhysRevLett.96.032501} {\bibfield  {journal} {\bibinfo  {journal} {Phys. Rev. Lett.}\ }\textbf {\bibinfo {volume} {96}},\ \bibinfo {pages} {032501} (\bibinfo {year} {2006})}\BibitemShut {NoStop}%
\bibitem [{\citenamefont {Clark}\ and\ \citenamefont {Macchiavelli}(2008)}]{Clark2008}%
  \BibitemOpen
  \bibfield  {author} {\bibinfo {author} {\bibfnamefont {R.~M.}\ \bibnamefont {Clark}}\ and\ \bibinfo {author} {\bibfnamefont {A.~O.}\ \bibnamefont {Macchiavelli}},\ }\bibfield  {title} {\bibinfo {title} {Exact and collective treatments of the pairing phase transition},\ }\href {https://doi.org/10.1103/physrevc.77.057301} {\bibfield  {journal} {\bibinfo  {journal} {Phys. Rev. C}\ }\textbf {\bibinfo {volume} {77}},\ \bibinfo {pages} {057301} (\bibinfo {year} {2008})}\BibitemShut {NoStop}%
\bibitem [{\citenamefont {Bès}\ \emph {et~al.}(1970)\citenamefont {Bès}, \citenamefont {Broglia}, \citenamefont {Perazzo},\ and\ \citenamefont {Kumar}}]{Bes1970}%
  \BibitemOpen
  \bibfield  {author} {\bibinfo {author} {\bibfnamefont {D.~R.}\ \bibnamefont {Bès}}, \bibinfo {author} {\bibfnamefont {R.~A.}\ \bibnamefont {Broglia}}, \bibinfo {author} {\bibfnamefont {R.~P.~J.}\ \bibnamefont {Perazzo}},\ and\ \bibinfo {author} {\bibfnamefont {K.}~\bibnamefont {Kumar}},\ }\bibfield  {title} {\bibinfo {title} {Collective treatment of the pairing {Hamiltonian}: ({I}). {Formulation} of the model},\ }\href {https://doi.org/10.1016/0375-9474(70)90677-9} {\bibfield  {journal} {\bibinfo  {journal} {Nucl. Phys. A}\ }\textbf {\bibinfo {volume} {143}},\ \bibinfo {pages} {1} (\bibinfo {year} {1970})}\BibitemShut {NoStop}%
\bibitem [{\citenamefont {Dussel}\ \emph {et~al.}(1971)\citenamefont {Dussel}, \citenamefont {Perazzo}, \citenamefont {Bès},\ and\ \citenamefont {Broglia}}]{Dussel1971}%
  \BibitemOpen
  \bibfield  {author} {\bibinfo {author} {\bibfnamefont {G.~G.}\ \bibnamefont {Dussel}}, \bibinfo {author} {\bibfnamefont {R.~P.~J.}\ \bibnamefont {Perazzo}}, \bibinfo {author} {\bibfnamefont {D.~R.}\ \bibnamefont {Bès}},\ and\ \bibinfo {author} {\bibfnamefont {R.~A.}\ \bibnamefont {Broglia}},\ }\bibfield  {title} {\bibinfo {title} {Collective treatment of the pairing {Hamiltonian}: ({II}). {Charge}-independent force -- {Hamiltonian} and symmetries},\ }\href {https://doi.org/10.1016/0375-9474(71)90446-5} {\bibfield  {journal} {\bibinfo  {journal} {Nucl. Phys. A}\ }\textbf {\bibinfo {volume} {175}},\ \bibinfo {pages} {513} (\bibinfo {year} {1971})}\BibitemShut {NoStop}%
\bibitem [{\citenamefont {Dussel}\ \emph {et~al.}(1972)\citenamefont {Dussel}, \citenamefont {Perazzo},\ and\ \citenamefont {Bès}}]{Dussel1972}%
  \BibitemOpen
  \bibfield  {author} {\bibinfo {author} {\bibfnamefont {G.~G.}\ \bibnamefont {Dussel}}, \bibinfo {author} {\bibfnamefont {R.~P.~J.}\ \bibnamefont {Perazzo}},\ and\ \bibinfo {author} {\bibfnamefont {D.~R.}\ \bibnamefont {Bès}},\ }\bibfield  {title} {\bibinfo {title} {Collective treatment of the pairing {Hamiltonian}: ({III}). {Numerical} solutions for the {$T=1$} case},\ }\href {https://doi.org/10.1016/0375-9474(72)90661-6} {\bibfield  {journal} {\bibinfo  {journal} {Nucl. Phys. A}\ }\textbf {\bibinfo {volume} {183}},\ \bibinfo {pages} {298} (\bibinfo {year} {1972})}\BibitemShut {NoStop}%
\bibitem [{\citenamefont {Bès}\ \emph {et~al.}(1973)\citenamefont {Bès}, \citenamefont {Dussel}, \citenamefont {Maqueda},\ and\ \citenamefont {Perazzo}}]{Bes1973}%
  \BibitemOpen
  \bibfield  {author} {\bibinfo {author} {\bibfnamefont {D.~R.}\ \bibnamefont {Bès}}, \bibinfo {author} {\bibfnamefont {G.~G.}\ \bibnamefont {Dussel}}, \bibinfo {author} {\bibfnamefont {E.~E.}\ \bibnamefont {Maqueda}},\ and\ \bibinfo {author} {\bibfnamefont {R.~P.~J.}\ \bibnamefont {Perazzo}},\ }\bibfield  {title} {\bibinfo {title} {Collective treatment of the pairing {Hamiltonian}: ({IV}). the yrast approximation},\ }\href {https://doi.org/10.1016/0375-9474(73)90625-8} {\bibfield  {journal} {\bibinfo  {journal} {Nucl. Phys. A}\ }\textbf {\bibinfo {volume} {217}},\ \bibinfo {pages} {93} (\bibinfo {year} {1973})}\BibitemShut {NoStop}%
\bibitem [{\citenamefont {Perazzo}\ \emph {et~al.}(1978)\citenamefont {Perazzo}, \citenamefont {Reich},\ and\ \citenamefont {Bès}}]{Perazzo1978}%
  \BibitemOpen
  \bibfield  {author} {\bibinfo {author} {\bibfnamefont {R.~P.~J.}\ \bibnamefont {Perazzo}}, \bibinfo {author} {\bibfnamefont {S.~L.}\ \bibnamefont {Reich}},\ and\ \bibinfo {author} {\bibfnamefont {D.~R.}\ \bibnamefont {Bès}},\ }\bibfield  {title} {\bibinfo {title} {Collective treatment of the pairing hamiltonian: ({V}). {Analysis} of transfer data for {$88 \le A \le 96$}},\ }\href {https://doi.org/10.1016/0375-9474(78)90511-0} {\bibfield  {journal} {\bibinfo  {journal} {Nucl. Phys. A}\ }\textbf {\bibinfo {volume} {311}},\ \bibinfo {pages} {219} (\bibinfo {year} {1978})}\BibitemShut {NoStop}%
\bibitem [{\citenamefont {Ni}\ and\ \citenamefont {Nakatsukasa}(2018)}]{PhysRevC.97.044310}%
  \BibitemOpen
  \bibfield  {author} {\bibinfo {author} {\bibfnamefont {F.}~\bibnamefont {Ni}}\ and\ \bibinfo {author} {\bibfnamefont {T.}~\bibnamefont {Nakatsukasa}},\ }\bibfield  {title} {\bibinfo {title} {Comparative study of the requantization of the time-dependent mean field for the dynamics of nuclear pairing},\ }\href {https://doi.org/10.1103/PhysRevC.97.044310} {\bibfield  {journal} {\bibinfo  {journal} {Phys. Rev. C}\ }\textbf {\bibinfo {volume} {97}},\ \bibinfo {pages} {044310} (\bibinfo {year} {2018})}\BibitemShut {NoStop}%
\bibitem [{\citenamefont {Ni}\ \emph {et~al.}(2018)\citenamefont {Ni}, \citenamefont {Hinohara},\ and\ \citenamefont {Nakatsukasa}}]{PhysRevC.98.064327}%
  \BibitemOpen
  \bibfield  {author} {\bibinfo {author} {\bibfnamefont {F.}~\bibnamefont {Ni}}, \bibinfo {author} {\bibfnamefont {N.}~\bibnamefont {Hinohara}},\ and\ \bibinfo {author} {\bibfnamefont {T.}~\bibnamefont {Nakatsukasa}},\ }\bibfield  {title} {\bibinfo {title} {Low-lying collective excited states in nonintegrable pairing models based on the stationary-phase approximation to the path integral},\ }\href {https://doi.org/10.1103/PhysRevC.98.064327} {\bibfield  {journal} {\bibinfo  {journal} {Phys. Rev. C}\ }\textbf {\bibinfo {volume} {98}},\ \bibinfo {pages} {064327} (\bibinfo {year} {2018})}\BibitemShut {NoStop}%
\end{thebibliography}%
\end{document}